\documentclass[lettersize,journal]{IEEEtran}
\usepackage{amsmath,amsfonts}
\usepackage{array}
\usepackage{tikz}
\usepackage{amsmath,mathtools}
\usepackage{amssymb}
\usepackage{booktabs}
\usepackage{graphicx}
\usepackage{amsmath}
\usepackage{algorithm}
\usepackage{algpseudocode}
\usepackage{boxedminipage}
\usepackage{fancyhdr}
\usepackage{multirow}
\usepackage{mathptmx}
\usepackage{array}
\usepackage{textcomp}
\usetikzlibrary{positioning}
\usetikzlibrary{arrows.meta}
\usetikzlibrary{calc}
\usetikzlibrary{shapes.geometric}
\usetikzlibrary{fit}
\usepackage{stfloats}
\usepackage{csquotes} 
\usepackage{url}
\usepackage{verbatim}
\usepackage{cite}
\usepackage{eqnarray}
\usepackage[caption=false,font=normalsize,labelfont=sf,textfont=sf]{subfig}
\usepackage{textcomp}
\usepackage{color}
\usepackage{stfloats}
\usepackage{url}
\usepackage{verbatim}
\usepackage{graphicx}
\def\BibTeX{{\rm B\kern-.05em{\sc i\kern-.025em b}\kern-.08em
    T\kern-.1667em\lower.7ex\hbox{E}\kern-.125emX}}
\usepackage{balance}
\usepackage{pifont}
\newcommand{\cmark}{\ding{51}}
\newcommand{\xmark}{\ding{55}}

\begin{document}
		\title{Semantic-Aware 6G Network Management through Knowledge-Defined Networking}

\author{
Tuğçe Bilen, \textit{Member, IEEE}, Ian F. Akyildiz, \textit{Life Fellow, IEEE}
\thanks{
Tuğçe Bilen is with the Department of Artificial Intelligence and Data Engineering, 
Faculty of Computer and Informatics, 
Istanbul Technical University, Istanbul, Turkey 
(e-mail: bilent@itu.edu.tr).}
\thanks{
Ian F. Akyildiz is with Truva Inc., Alpharetta, GA 30022, USA 
(e-mail: ian@truvainc.com).}
}

\markboth{Submitted To Transactions on Mobile Computings,~Vol.~xx, No.~xx, February~2026}%
{How to Use the IEEEtran \LaTeX \ Templates}

\maketitle

\begin{abstract}
	Semantic communication is emerging as a key paradigm for 6G networks, where the goal is not to perfectly reconstruct bits but to preserve the meaning that matters for a given task. This shift can improve bandwidth efficiency, robustness, and application-level performance. However, most existing studies focus solely on encoder-decoder design and ignore network-wide decision-making. As data traverses multiple hops, semantic relevance may decrease, routing may overlook meaningful information, and semantic distortion can increase under dynamic network conditions. To address these challenges, this paper proposes a management-oriented semantic communication framework built upon Knowledge-Defined Networking (KDN). The framework comprises three core modules: a semantic-reasoning module that computes relevance scores by mapping semantic embeddings onto a knowledge graph that encodes task concepts and contextual relationships; a semantic-aware routing mechanism that forwards data along paths that preserve meaning; and a semantic-distortion controller that adaptively adjusts encoding and routing to preserve semantic fidelity. Our ns-3 results show clear benefits: semantic delivery success improves by 12\%, semantic distortion decreases by 22\%, re-routing events drop by 44\%, and throughput efficiency rises by 14\% compared to baseline methods (shortest-path, load-based, and distortion-only routing). These results indicate that meaning-aware and feedback-driven control is essential for reliable and scalable semantic communication in future 6G networks.
\end{abstract}


	\section{Introduction}
	Sixth-generation (6G) wireless networks will support a broad range of applications, including autonomous transportation, real-time digital twins, large-scale IoT, and immersive extended reality \cite{9598915}. These services require strict performance guarantees, including low latency, high reliability, and efficient use of communication and computation resources. 6G must meet these different requirements, which go beyond traditional bit-level optimization. This drives the use of intelligent methods that make transmitted information more efficient and meaningful.
	
	Semantic communication has recently gained attention as a promising approach for 6G systems \cite{liu2024survey}. Instead of focusing on the accurate reconstruction of bits, semantic communication aims to preserve the meaning that is relevant to the task at the receiver. By transmitting only task-related information, semantic communication can reduce bandwidth consumption, improve robustness under variable network conditions, and support applications that rely on real-time decision-making. These characteristics make semantic communication well suited to 6G environments, where network conditions change rapidly and application intent shapes communication requirements. 
	
\begin{figure}[h]
		\centering
		\resizebox{\linewidth}{!}{%
			
			\begin{tikzpicture}[
				font=\sffamily,
				node distance=1.2cm and 2.5cm,
				box/.style={
					rectangle,
					rounded corners,
					draw,
					minimum width=5cm,
					minimum height=1.1cm,
					align=center
				},
				smallbox/.style={
					rectangle,
					rounded corners,
					draw,
					minimum width=4.7cm,
					minimum height=1.4cm,
					align=center,
					text width=4.7cm
				},
				contrib/.style={
					rectangle,
					rounded corners,
					draw,
					fill=gray!5,
					minimum width=4.7cm,
					minimum height=1.4cm,
					align=center,
					text width=4.7cm
				},
				framework/.style={
					fill=gray!15,
					rectangle,
					rounded corners,
					draw,
					minimum width=16cm,
					minimum height=4.5cm,
					align=center
				},
				>={Stealth[length=2mm,width=2mm]}
				]
				
				
				\node[box] (sixg) {Sixth-Generation (6G) Networks};
				\node[box, below=1.1cm of sixg] (sem) {Semantic Communication};
				
				
				\node[smallbox, below left=1.7cm and 3.8cm of sem] (c1)
				{    Semantic Representations\\
					\footnotesize Meaning may diminish over multi-hop paths};
				
				\node[smallbox, below=1.7cm of sem] (c2)
				{
					Routing Decisions\\
					\footnotesize May ignore semantic importance};
				
				\node[smallbox, below right=1.7cm and 3.8cm of sem] (c3)
				{
					Semantic Distortion\\
					\footnotesize Meaning may drift due to fluctuating network states};
				
				
				\node[framework, below=1.0cm of c2] (fw) {};
				
				\node at ($(fw.south)+(0,0.7)$)
				{\large\bf Proposed KDN-Driven Semantic Communication Framework};
				
				
				\node[contrib, above=1.4cm of fw.south, xshift=-5.3cm] (s1)
				{Semantic Reasoning Module\\
					\footnotesize Computes semantic relevance};
				
				\node[contrib, above=1.4cm of fw.south] (s2)
				{Semantic-Aware Routing\\
					\footnotesize Uses relevance-weighted metrics};
				
				\node[contrib, above=1.4cm of fw.south, xshift=5.3cm] (s3)
				{Semantic Distortion Control\\
					\footnotesize Preserves semantic fidelity};
				
				
				\draw[->, thick] (sixg) -- (sem);
				
				\draw[->, thick] (sem) -- (c1);
				\draw[->, thick] (sem) -- (c2);
				\draw[->, thick] (sem) -- (c3);
				
				\draw[->, thick] (c1.south) -- ++(0,-0.6) -| (s1.north);
				\draw[->, thick] (c2.south) -- (s2.north);
				\draw[->, thick] (c3.south) -- ++(0,-0.6) -| (s3.north);
				
			\end{tikzpicture}
			
		}%
		\caption{Challenges and corresponding modules within the proposed semantic communication framework.}
		\label{fig:challenge-solution-map}
	\end{figure}
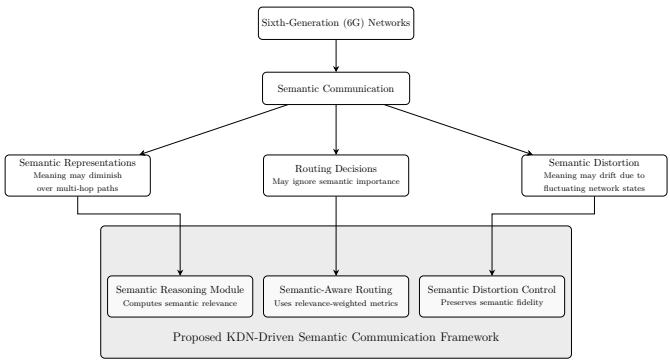

	Although semantic communication offers clear advantages, several challenges remain in integrating it into large-scale 6G networks. Most existing semantic methods focus on designing encoder–decoder models and assume stable channels or fixed communication patterns. In practice, 6G networks comprise multi-hop paths, heterogeneous nodes, mobility, and dynamic service demands. Under these conditions, semantic representations may become less relevant, routing decisions may ignore semantic importance, and semantic distortion may increase due to network variations, as summarized in Fig. \ref{fig:challenge-solution-map}. Moreover, current systems lack a network-level mechanism that reasons about semantic importance, prioritizes meaning-critical information, and provides feedback to preserve semantic consistency during end-to-end transmission.

        Crucially, existing semantic communication models cannot be simply extended to the networking layer because they are fundamentally local optimizations. These models treat semantic encoding and decoding as isolated operations at the endpoints, assuming that meaning degradation occurs only along a single link and can be fully compensated for by better codec design. However, in multi-hop 6G networks, semantic distortion accumulates non-linearly across paths, depends on dynamic interactions between flows, and varies with routing decisions that are themselves unaware of semantic content. Extending endpoint-centric semantic models to the network level would require them to predict and correct for congestion dynamics, mobility-induced topology changes, and cross-traffic interference. This fundamental mismatch between local semantic codec design and global network dynamics necessitates a new architectural approach where semantics becomes a first-class control variable within the network management plane itself.

As illustrated in Fig. \ref{fig:challenge-solution-map}, this paper addresses these challenges by introducing a Knowledge-Defined Networking (KDN)-driven semantic communication framework that supports semantic reasoning, semantic-aware routing, and semantic distortion control in 6G networks. 
\color{black}{Unlike prior works that treat these functions in isolation or focus primarily on encoder--decoder optimization, the proposed framework unifies them into a coordinated, network-wide control architecture.} \color{black} Our approach treats semantics as a network-level attribute that can guide communication decisions beyond basic throughput or delay. 
\color{black}{In this sense, the contribution is not a new standalone algorithm, but a system-level closed-loop semantic management paradigm for multi-hop 6G networks.} \color{black} The goal is to help 6G networks identify relevant information, forward it over suitable paths, and preserve semantic fidelity under dynamic conditions. The key contributions of this paper are listed as follows:
\begin{itemize}

\item We design a semantic reasoning module that computes relevance scores from contextual information, task requirements, and concept-level representations, enabling explicit network-level semantic decision making rather than application-local filtering. These scores guide adaptive semantic encoding and interpretation.

\item We propose a semantic-aware routing mechanism that forwards data according to semantic importance. Unlike traffic- or delay-driven routing, it jointly considers semantic relevance and end-to-end distortion, allowing the network to allocate resources based on task needs instead of bit-level metrics.

\item We develop a closed-loop semantic distortion management process that monitors end-to-end semantic consistency and links semantic degradation to both routing and encoding decisions. This feedback loop dynamically adjusts encoding fidelity and routing under mobility and changing network conditions.

\item We integrate these components within a KDN architecture, transforming the Knowledge Plane into a semantic control entity that coordinates reasoning, routing, and correction at network scale, while the Data Plane provides real-time semantic feedback.

\item We propose an evaluation methodology that measures semantic fidelity, routing efficiency, and robustness across diverse network conditions, targeting system-level semantic management behavior rather than isolated link effects and remaining scalable without detailed physical-layer modeling.

\end{itemize}
\color{black}
	
The remainder of the paper is organized as follows. Section II reviews related work. Section III presents the problem formulation and system architecture. Section IV describes the proposed approach, while Section V evaluates its performance through simulations. Section VI concludes the paper, and Section VII discusses future research directions.

	\section{Related Work}	
	\subsection{Semantic Communication Models}
	Most semantic communication studies focus on learning-based encoder--decoder design. DeepSC, introduced in \cite{9398576}, is one of the first deep learning semantic communication systems for text transmission. It uses Transformer-based semantic extraction and joint source--channel coding to preserve meaning under noisy channels. Later works extend this model to other modalities and tasks. For example, DeepSC-ST \cite{10038754} develops a semantic encoder--decoder pipeline for speech transmission by directly aligning semantic features with speech recognition and synthesis tasks. In parallel, deep joint source--channel coding frameworks for semantics have been studied in \cite{10328187}, where semantic representations are compressed and transmitted through adaptive neural architectures. Recent studies, such as \cite{Qin2021SemanticCP}, further formalise semantic communication principles and outline model-level challenges, including semantic noise, context mismatch, and task coupling. Although these works demonstrate strong semantic compression and robustness, they are mostly limited to single-link or simplified scenarios. Their optimization is centered on encoder--decoder fidelity, without a network-level mechanism for semantic reasoning, semantic-aware routing, or semantic distortion control. Thus, meaning extraction is optimized locally, but end-to-end semantic management over multi-hop 6G networks remains open.
    
\begin{table*}[t]
\centering
\caption{Capabilities of Major Research Directions}
\label{tab:related-work-summary}
\renewcommand{\arraystretch}{1}
\scriptsize
\begin{tabular}{c|c|c|c|c}
\toprule
\textbf{Research Direction} &
\textbf{Semantic Reasoning} &
\textbf{Semantic-Aware Routing} &
\textbf{Distortion Control} &
\textbf{KDN-Based Management} \\
\hline\hline

Semantic Communication Models 
{\cite{9398576,10038754,10328187,Qin2021SemanticCP}}
& Partial & $\times$ & $\times$ & $\times$ \\

\hline

Semantic-Aware Networking / Task-Oriented
{\cite{DBLP:journals/corr/abs-2012-15405,9955525,9606667,10183786}}
& $\checkmark$ & $\times$ & $\times$ & $\times$ \\

\hline

Semantic Distortion and Fidelity
{\cite{10122224,10110357,10454584}}
& $\checkmark$ & $\times$ & $\checkmark$ & $\times$ \\

\hline

KDN for Network Management
{\cite{10.1145/3138808.3138810,ASHTARI2022100136,telecom4030025,10012574}}
& $\times$ & $\checkmark$ & $\times$ & $\checkmark$ \\

\hline

\textbf{This Work}
& $\checkmark$ & $\checkmark$ & $\checkmark$ & $\checkmark$ \\

\bottomrule
\end{tabular}
\end{table*}

	\subsection{Semantic-Aware Networking}
	A second research line shifts attention from model design to semantic relevance at the networking level. The concept of semantic-aware networking is introduced in \cite{DBLP:journals/corr/abs-2012-15405}, where the authors argue that future networks should exploit meaning and task intent to guide networking functions. Task-oriented communication is further developed in \cite{9955525}, which positions semantics and context as central elements of next-generation wireless systems. Beyond conceptual frameworks, task-oriented learning-based communication is studied in \cite{9606667}, where information-bottleneck methods compress task-relevant features for edge inference. Resource-level adaptation for task-oriented semantic systems is further explored in \cite{10183786}, showing that semantic importance can guide spectrum and compute allocation. These studies still do not provide operational multi-hop semantic routing or closed-loop distortion mitigation. Semantic relevance is usually handled at the application or link level, while routing decisions remain meaning-agnostic. Therefore, a network management framework that explicitly routes traffic based on semantic importance and preserves semantic fidelity across paths remains missing.

	\subsection{Semantic Distortion and Fidelity Studies}
	Semantic distortion has recently been formalized as a key metric for semantic communication. Authors provide a rate--distortion interpretation of goal-oriented semantic communication and define fidelity constraints that quantify meaning loss in \cite{10122224}. In a similar direction, explainable task-oriented semantic systems \cite{10110357} analyze semantic degradation under semantic noise and measure semantic accuracy using embedding-based similarity. More broadly, semantic distortion is treated as a native networking concern in \cite{10454584}, which discusses semantic entropy, semantic mutual information, and fidelity-aware communication objectives. Despite these advances, prior work mainly measures semantic distortion rather than managing it. Distortion is computed after decoding but is not used as feedback to adapt routing, semantic policies, or encoding fidelity at the network level. This motivates our closed-loop semantic distortion control module embedded into a KDN-based management plane.
	
	\subsection{KDN for 5G/6G Management}
	KDN defines an architectural paradigm that integrates SDN, telemetry, and machine learning into a dedicated Knowledge Plane \cite{BILEN2025103984}, \cite{11357882}, \cite{BILEN2026111941}. The KDN architecture and its reasoning-driven operation are presented in \cite{10.1145/3138808.3138810}. Subsequent surveys such as \cite{ASHTARI2022100136} and \cite{telecom4030025} summarize KDN use cases in traffic prediction, routing optimization, anomaly detection, and resource control. Concrete AI-driven routing solutions in KDN environments are studied in \cite{10012574}. According to it, a Knowledge Plane can learn routing policies based on global network state. However, KDN literature has so far focused on bit-level KPIs such as delay, throughput, and congestion. None of these studies incorporates semantic importance, semantic relevance reasoning, or distortion feedback as first-class control inputs. 
Therefore, integrating semantic communication into KDN-based 6G management remains largely unexplored and constitutes the paper's core novelty.

As summarized in Table~\ref{tab:related-work-summary}, across these four research threads, prior work either (i) focuses solely on semantic encoders and decoders, (ii) studies semantic intent at a high conceptual level without operational mechanisms, (iii) measures semantic distortion without actively managing it, or (iv) applies KDN to traditional networking objectives. 
\color{black}{In contrast, our framework introduces a coordinated, closed-loop semantic management plane in which semantic reasoning, relevance-aware routing, and distortion-driven adaptation are jointly embedded into the network control loop. By treating semantics as a first-class, network-level control signal, the proposed architecture enables continuous semantic decision making across multi-hop paths under dynamic load and mobility conditions.} \color{black}
To the best of our knowledge, no study has combined semantic reasoning, semantic-aware routing, and closed-loop distortion control within a KDN-based architecture for 6G network management, a gap this paper addresses.

	\section{The Proposed System Model}
    This section begins by formalizing the core problem addressed in this study, followed by a detailed presentation of the proposed system model.
\subsection{Problem Statement}
Semantic communication enables 6G networks to transmit task-relevant meaning rather than raw data. However, deploying semantic communication in a dynamic multi-hop network introduces several management challenges. Semantic representations may lose relevance as network conditions change, routing decisions may ignore message importance, and semantic distortion may increase under mobility or congestion. To operate reliably, the network must recognize the importance of semantic messages, select appropriate forwarding paths, and preserve meaning under varying conditions.

We consider a 6G network where a message $m$ is generated at a source node and encoded into a semantic vector $\mathbf{s}(m)$. The message is delivered to a destination through a multi-hop path 
\textcolor{black}{$p \in \mathcal{P}$, where $\mathcal{P}$ denotes the candidate path set and the finally selected path is denoted by $p^\star(m)$.}
Each message is associated with a semantic relevance score $R(m)$, an encoding fidelity level 
\textcolor{black}{$f \in \mathcal{F}$, where the selected fidelity is denoted by $f^\star(m)$,}
and an expected semantic distortion value 
\textcolor{black}{$\hat{D}(m,p,f)$}
that depends on both the selected path, the chosen fidelity level, and the current network state. Different paths in $\mathcal{P}$ have different delay, load, mobility, and expected distortion characteristics. The Data Plane forwards packets along the selected path, while the Knowledge Plane monitors network metrics and makes routing and adaptation decisions. The goal is to minimize semantic distortion under resource constraints, as defined in Eq.~\ref{1}.

\small
\begin{equation}\label{1}
\begin{aligned}
&\min_{p \in \mathcal{P},\, f \in \mathcal{F}} \; \hat{D}(m,p,f) \\
\text{s.t.}\quad 
&\mathrm{delay}(p) \le \Delta_{\max}, \\
&\mathrm{load}(p) \le L_{\max}.
\end{aligned}
\end{equation}

\normalsize

\textcolor{black}{Here, $\Delta_{\max}$ denotes the maximum tolerable end-to-end delay for the message, reflecting application latency requirements, and $L_{\max}$ denotes the maximum admissible path load to avoid excessive congestion and instability.} Also, $\mathcal{F}$ is the set of allowable fidelity levels. Higher fidelity reduces distortion but increases bandwidth and processing cost, while lower fidelity saves resources at the expense of meaning accuracy. \textcolor{black}{The solution of the optimization problem yields the optimal path $p^\star(m)$ and the optimal fidelity level $f^\star(m)$, which are subsequently enforced by the routing and encoding modules.} \textcolor{black}{In addition to the predicted distortion $\hat{D}(m,p,f)$ used for decision making, the system also measures the \emph{observed semantic distortion} $D_{\text{obs}}(m)$ after delivery, which is computed at the receiver by comparing the reconstructed semantic vector with the original $\mathbf{s}(m)$. This feedback is used by the distortion control loop to validate and adjust routing or fidelity decisions.} Based on these conditions, the following three core challenges arise:

\begin{itemize}
	\item {Semantic relevance identification:} This phase computes $R(m)$ using task information, context, and semantic relationships.
	\item {Semantic-aware routing:} This phase selects a path 
	\textcolor{black}{$p^\star(m) \in \mathcal{P}$}
	that matches message importance to ensure reliable and low-distortion delivery.
	\item {Semantic distortion management:} This phase monitors distortion and adapts fidelity or routing when 
	\textcolor{black}{$\hat{D}(m,p,f)$ predicted by the model or the observed distortion $D_{\text{obs}}(m)$}
	exceed acceptable limits.
\end{itemize}

Solving these challenges requires a management framework that links semantic reasoning, adaptive routing, and distortion control. The next subsection presents the system architecture used to implement these functionalities within a KDN-based framework.

	\subsection{System Model}
This work proposes a management-oriented semantic communication architecture for 6G networks. The architecture separates data forwarding, semantic processing, and network-level decision-making into distinct modules. This separation allows each function to evolve independently while still operating in a tightly coordinated manner. 
\color{black}{Also, different from application-centric semantic communication models, this design explicitly treats semantics as a control variable at the network level.} \color{black} The proposed system consists of two main components: the Data Plane and the Knowledge Plane. The Data Plane handles semantic message generation, transmission, and telemetry collection, whereas the Knowledge Plane performs high-level control decisions based on real-time semantic and network context. \color{black}{Together, these planes form a closed control loop in which semantic interpretation, forwarding, and correction are jointly optimized at the system scale.} \color{black} These components exchange information continuously to support semantic reasoning, semantic-aware routing, and semantic distortion control throughout the entire communication process. The details are explained as follows:

 \color{black}
\subsubsection{Data Plane} 
The Data Plane contains the operational nodes of the 6G network, including base stations, access points, edge servers, and relays. Each node runs a lightweight semantic encoder that transforms raw data into a semantic representation vector $\mathbf{s}(m)$ capturing entities, intent, and contextual meaning. The corresponding decoder reconstructs this semantic representation at downstream nodes or at the destination.
\color{black}{These encoding and decoding functions are intentionally kept lightweight, as semantic intelligence and policy decisions are handled in the Knowledge Plane rather than at individual forwarding nodes.}  \color{black} Each node also collects real-time network measurements that influence semantic quality, including delay, loss rate, queue occupancy, mobility indicators, and traffic load. These metrics are periodically reported to the Knowledge Plane as a network state vector 
\color{black}{$\mathbf{n}(t)$. This vector aggregates time-varying link- and node-level measurements for global decision-making.} \color{black}
In addition, nodes estimate semantic distortion by comparing the transmitted semantic representation with its locally reconstructed version, producing an observed distortion value $D_{\text{obs}}(m)$. 
\color{black}{This local estimation serves as a measurement signal rather than a decision variable, and is forwarded to the Knowledge Plane to support distortion monitoring and control.}\color{black}
This feedback indicates the extent to which the message's meaning is preserved along the current forwarding path. During packet forwarding, Data Plane nodes do not perform semantic reasoning. 
\color{black}{They execute forwarding and encoding actions according to the path $p^\star(m)$ and fidelity level $f^\star(m)$ determined by the Knowledge Plane.} \color{black}
Routing and adaptation decisions are issued by the Knowledge Plane to keep the operational layer simple and enable central network-wide semantic intelligence.

\subsubsection{Knowledge Plane} 
The Knowledge Plane serves as the intelligent control layer. It receives telemetry and semantic-distortion feedback from the Data Plane and performs three primary tasks. First, it conducts \emph{semantic reasoning}. The Knowledge Plane maintains a knowledge graph containing task concepts and semantic dependencies. Using this graph, each message is assigned a relevance score $R(m)$ reflecting its importance to the current task. 
\color{black}{This relevance assessment is used as an explicit control signal rather than a passive annotation.}  
\color{black} This score influences the required encoding fidelity and routing priority. Second, it performs \emph{semantic-aware routing}. Rather than treating all packets equally, the Knowledge Plane selects forwarding paths based on semantic importance and network conditions. 
\color{black}{The output of this stage is the selected path $p^\star(m) \in \mathcal{P}$. It is determined using relevance-aware cost evaluation over the candidate path set.}  \color{black} High-relevance messages are forwarded through paths with stable delay and low expected distortion, while lower-priority messages may use best-effort routes to conserve resources. Third, it manages \emph{semantic distortion control}. The Knowledge Plane continuously compares the predicted semantic distortion 
\color{black}{$\hat{D}(m,p,f)$ evaluated for the current operational decisions}
with the observed distortion $D_{\text{obs}}(m)$ reported by the Data Plane. \color{black}{After a path $p^\star(m)$ and fidelity level $f^\star(m)$ have been selected and applied, the observed distortion $D_{\text{obs}}(m)$ reflects the actual semantic degradation under those decisions.}
If the observed distortion exceeds the relevance-dependent tolerance threshold $\delta(m)$, the Knowledge Plane updates the encoding fidelity level $f^\star(m)$, triggers re-routing by recomputing $p^\star(m)$, or adjusts semantic control parameters.  
\color{black} \color{black}{The output of this stage is therefore an updated control decision $(p^\star(m), f^\star(m))$ when correction is required.}
This mechanism closes the semantic control loop by linking observed degradation to corrective network actions.  
\color{black}
This loop preserves semantic quality under dynamic conditions, including mobility, congestion, and link variability.

Finally, a continuous feedback loop links both planes. The Data Plane sends real-time state information and distortion measurements, and the Knowledge Plane returns updated routing and adaptation commands $(p^\star(m), f^\star(m))$. This modular architecture supports scalability and flexibility. New reasoning or routing modules can be added without modifying the underlying network, and the system adapts naturally to topology changes and mobility. 
\color{black}{In this sense, the architecture should be understood as a network-wide semantic management framework rather than a collection of independent optimization functions.}

	\begin{figure*}[h]
		\centering
		\includegraphics[width=0.8\textwidth]{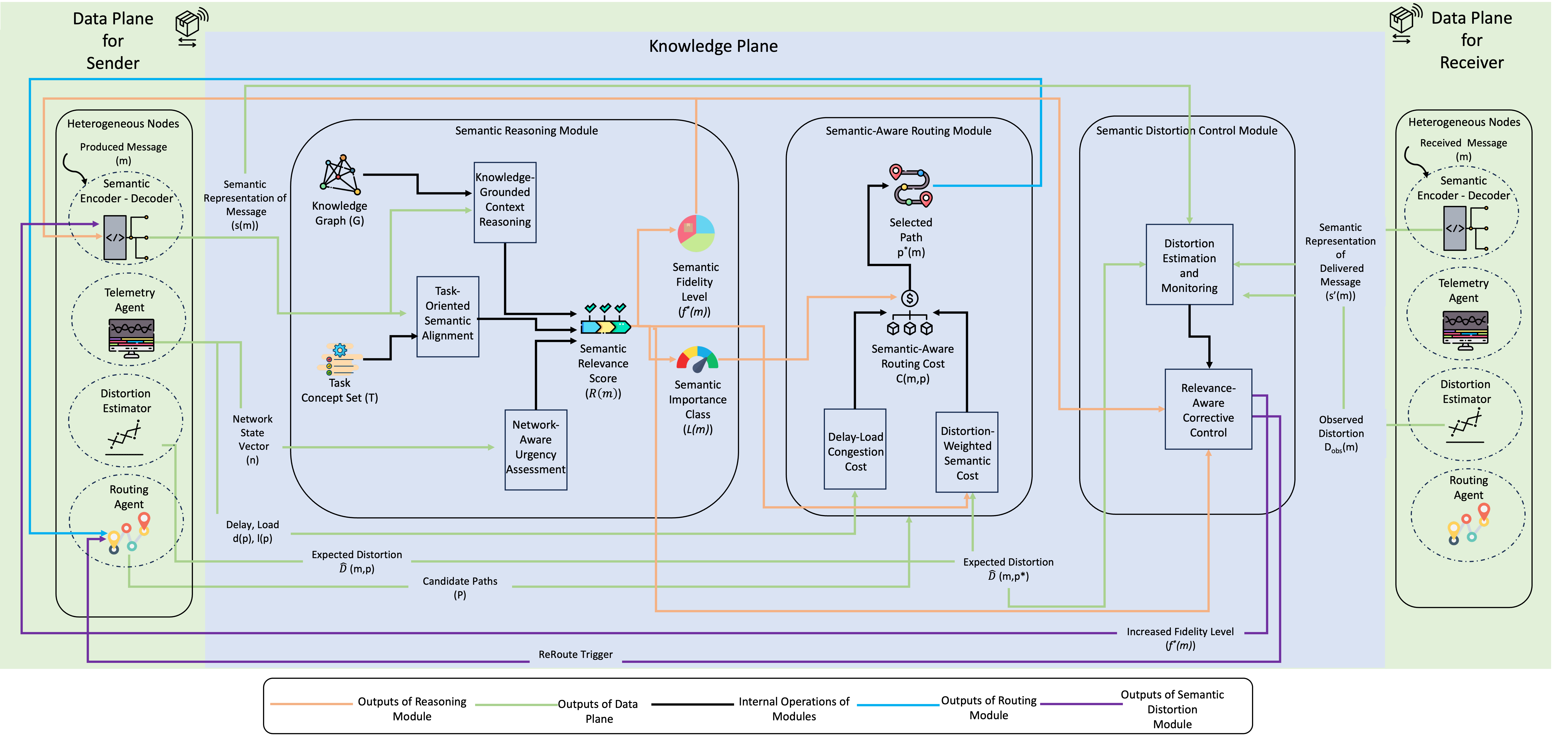}	
		\caption{The proposed framework}
		\label{model}
	\end{figure*}

	\section{The Proposed Framework}
	This section presents the proposed management framework that enables semantic communication in 6G networks. As illustrated in Fig. \ref{model}, the framework integrates three core mechanisms: (i) semantic reasoning, (ii) semantic-aware routing, and (iii) semantic distortion control. Each mechanism is implemented in the Knowledge Plane and interacts with the Data Plane through a feedback loop.

\subsection{Semantic Reasoning Module}
The semantic reasoning module assesses the importance of each semantic message to the ongoing task at the destination. This score drives both the semantic-aware 
routing and distortion control mechanisms. The module processes each incoming message using four inputs: its semantic representation $\mathbf{s}(m)$, the task concept set $\mathcal{T}$, the contextual relations stored in the knowledge graph $\mathcal{G}$, and the instantaneous network state vector $\mathbf{n}$.

When a message $m$ is generated at a source node, the semantic encoder in the Data Plane produces a semantic representation vector $\mathbf{s}(m)$ describing 
the entities, intent, and contextual meaning of the message. This vector is forwarded to the Knowledge Plane for reasoning. In parallel, each Data Plane node 
runs a lightweight telemetry agent that continuously reports a network state vector $\mathbf{n}$ containing delay, queue occupancy, mobility indicators, loss 
rate, and traffic load. This allows the module to incorporate real-time network conditions into the relevance evaluation.

The Knowledge Plane also maintains two additional sources of background information. The first is a domain knowledge graph $\mathcal{G}$, which stores 
high-level relations between concepts (e.g., entity links, causal dependencies, or semantic associations). This graph is constructed offline from operator 
ontologies and application concept models and is maintained as a persistent knowledge base. 
\color{black}{The graph is not modified at packet time scale; instead, it represents relatively stable semantic structure used for contextual grounding in the reasoning process.}
The second is the task concept set $\mathcal{T}$ provided by the 
destination application, which specifies the semantic elements required for correct task execution and is communicated directly to the Knowledge Plane through an application interface.

\color{black}
The task concept set represents high-level task intent rather than instantaneous packet-level objectives and is therefore assumed to evolve on a slower timescale than per-message forwarding or routing decisions. Temporary mismatches, missing concepts, or coarse granularity in $\mathcal{T}$ do not immediately invalidate relevance estimation. Instead, their impact is moderated through continuous similarity-based scoring, contextual grounding over $\mathcal{G}$, and temporal normalization of relevance values before they are consumed by routing and distortion control modules. As a result, relevance scores evolve smoothly even under imperfect or gradually changing task specifications, preventing abrupt routing fluctuations.
\color{black}

\color{black}
Conceptually, the semantic reasoning module operates in three successive stages: (i) task-oriented semantic alignment between $\mathbf{s}(m)$ and $\mathcal{T}$, (ii) knowledge-grounded contextual reasoning using $\mathcal{G}$, and (iii) network-aware urgency assessment based on $\mathbf{n}$. These stages progressively refine the importance of a message, from task semantics to contextual validity, and finally to delivery urgency, under current network conditions.
\color{black}

Together, the inputs $\mathbf{s}(m)$, $\mathcal{G}$, $\mathcal{T}$, and $\mathbf{n}$ form the complete information set required by the semantic reasoning module, as summarized in Algorithm~\ref{alg:reasoning-flow}. Using these four inputs, the relevance of message $m$ is computed in the following three consecutive stages.

\subsubsection{Task-Oriented Semantic Alignment}
The first stage determines the degree to which the semantic content of message $m$ contributes to the execution of the target task at the receiver. In practical 6G scenarios, only a subset of the transmitted data contains information that is truly relevant to decision-making. Therefore, this stage prioritizes messages whose semantic embedding $\mathbf{s}(m)$ is aligned with the task concept set $\mathcal{T}$, which represents the knowledge required by the destination application.

To quantify this alignment, a similarity score $S(\mathbf{s}(m),\mathcal{T})$ is computed using cosine similarity between the semantic vector and each task concept vector in $\mathcal{T}$, as given in Eq.~\ref{2}. This formulation ensures that relevance depends on semantic meaning rather than low-level bit patterns or lexical resemblance.
\color{black}
Since $\mathcal{T}$ reflects high-level task intent rather than exact packet semantics, its elements are typically coarse-grained and evolve at the application time scale. Consequently, alignment is evaluated at the semantic-task level instead of relying on strict concept matching.
\color{black}

\small
\begin{equation}\label{2}
	S(\mathbf{s}(m),\mathcal{T}) =
	\max_{t \in \mathcal{T}}
	\frac{\mathbf{s}(m)\cdot \mathbf{t}}
	{|\mathbf{s}(m)|\,|\mathbf{t}|}
\end{equation}

\normalsize

Selecting the maximum similarity identifies the closest conceptual match within the task. This enforces a best-fit semantic alignment rule that preserves significant information while suppressing irrelevant or redundant messages. The use of normalized cosine similarity also improves robustness against representation shift or amplitude variations introduced by lossy transforms in the transmission chain.
\color{black}
If $\mathcal{T}$ contains missing, noisy, or partially outdated concepts, the max-similarity operation ensures that relevance decreases smoothly rather than collapsing abruptly, provided that at least one task-related concept remains semantically close to $\mathbf{s}(m)$. This behavior prevents transient task-concept inconsistencies from triggering unstable routing or distortion-control reactions.
\color{black} Since the complexity of the alignment step grows linearly with the number of task concepts, it is computationally tractable for real-time operation even under dense message arrivals.
\color{black}
Because updates to $\mathcal{T}$ occur at a slower time scale, their impact on $S(\mathbf{s}(m),\mathcal{T})$ propagates gradually across messages. Downstream modules therefore adapt through statistical normalization of relevance values rather than reacting to individual concept-level fluctuations. \color{black} Therefore, this stage acts as a task-driven semantic filter that prioritizes messages carrying information critical to the receiver.

\subsubsection{Knowledge-Grounded Context Reasoning}

The second stage refines the relevance estimation by using contextual information stored in the knowledge graph $\mathcal{G}$. Unlike the first stage, which checks direct similarity with task concepts, this step captures implicit meaning through related entities and semantic relationships.

First, the semantic vector $\mathbf{s}(m)$ is mapped to one or more concept nodes in $\mathcal{G}$ that represent the entities or attributes mentioned in the message. Then, a neighborhood set $\mathcal{N}(m)$ is formed by collecting the concept nodes directly connected to these mapped nodes through semantic relations such as parent--child links, attribute associations, or causal dependencies. In this way, $\mathcal{N}(m)$ represents the immediate semantic context of the message.
\color{black}
This graph-based expansion allows the reasoning process to remain effective even when the task concept set $\mathcal{T}$ is incomplete, coarsely defined, or temporarily outdated. Semantically related concepts can still contribute to relevance through their connections in the graph.
\color{black} The contextual consistency score is then computed as in Eq.~\ref{3}, where $w_{m,v}$ denotes the semantic affinity between message $m$ and its neighboring concept $v$.

\begin{algorithm}[h]
\caption{Semantic Reasoning Module}
\scriptsize
\label{alg:reasoning-flow}
\begin{algorithmic}[1]
	\Require semantic vector $\mathbf{s}(m)$, task concept set $\mathcal{T}$, 
	knowledge graph $\mathcal{G}$, network state vector $\mathbf{n}$
	
	\State Compute task-alignment score $S(\mathbf{s}(m), \mathcal{T})$ using Eq.~\ref{2}
	\State Compute knowledge-grounded context score $C(m)$ using Eq.~\ref{3}
	\State Map network state to urgency score $U(m) = \phi(\mathbf{n})$
	
	\State Fuse the three components into semantic relevance $R(m)$ using Eq.~\ref{4}
	\State Standardize relevance value $z(m)$ using Eq.~\ref{e1}
	\State Determine discrete importance class $L(m)$ using Eq.~\ref{e2}
	\State Select optimal fidelity level $f^\star(m)$ using Eq.~\ref{6}
	
	\State \Return $R(m)$, $L(m)$, $f^\star(m)$
\end{algorithmic}
\end{algorithm}

\small

\begin{equation} \label{3}
	C(m) =
	\frac{1}{|\mathcal{N}(m)|}
	\sum_{v \in \mathcal{N}(m)} w_{m,v}
\end{equation}

\normalsize

A higher $C(m)$ value indicates that the message is not only aligned with task-related concepts but also fits well within its broader semantic context in the knowledge graph. This improves robustness when semantic vectors are partially degraded or noisy, since contextual relations can still preserve meaning.
\color{black}
Errors or drift in the task concept set $\mathcal{T}$ therefore influence $C(m)$ only indirectly and gradually. Contextual relevance is inferred from relatively stable graph relations instead of exact task-concept membership. This prevents sudden changes in relevance and reduces the risk of routing or control instability caused by temporary task-level inconsistencies.
\color{black}

\subsubsection{Network-Aware Urgency Assessment}
The third stage evaluates how current network conditions affect the practical usefulness of delivering message $m$. While the first two stages capture task relevance and contextual meaning, a message may still lose value if delay, congestion, mobility, or link instability prevents timely and reliable delivery. Therefore, the reasoning module incorporates the instantaneous network state vector $\mathbf{n}$ to estimate delivery urgency. The vector $\mathbf{n}$ includes normalized indicators as end-to-end delay, queue occupancy, short-term load, link quality, and mobility level. These metrics are mapped to a scalar urgency score $U(m) = \phi(\mathbf{n})$, where $\phi(\cdot)$ is a monotonic function that assigns higher urgency under worsening network conditions. \color{black}
Thus, urgency is computed independently of the task concept set $\mathcal{T}$. This separation allows the control loop to respond to transient network degradations even when task-level relevance estimates are temporarily imprecise.
\color{black} As a result, messages with moderate semantic relevance may receive temporary priority when network dynamics threaten their successful delivery. This stage ensures that semantic value is evaluated jointly with delivery feasibility rather than in isolation.
\color{black}
To avoid instability, urgency updates rely on smoothed telemetry and bounded control intervals. This design reduces abrupt priority shifts and mitigates oscillatory behavior under fast mobility or bursty load variations.
\color{black}

These three components are linearly combined to form the semantic relevance score $R(m)$ in Eq.~\ref{4}. The weighting parameters 
$\alpha$, $\beta$, and $\gamma$ control the contribution of task alignment, semantic context, and urgency, respectively, 
and satisfy $\alpha + \beta + \gamma = 1$. A larger value of $R(m)$ indicates that preserving the meaning of message $m$ should be prioritized during transmission.
\color{black}
This linear fusion allows task semantics and short-term network dynamics to jointly influence downstream control decisions while preserving stability across control intervals.
\color{black}

\small
\begin{equation} \label{4}
	R(m) = \alpha\, S(\mathbf{s}(m),\mathcal{T})
	+ \beta\, C(m)
	+ \gamma\, U(m)
\end{equation}
\normalsize

After computing the continuous relevance score, the message is normalized to capture its relative importance within recent traffic. Let $\mu_R$ and $\sigma_R$ denote the running mean and standard deviation of recent relevance values. The standardized score is given in Eq. \ref{e1}.

\small
\begin{equation} \label{e1}
	z(m) = \frac{R(m) - \mu_R}{\sigma_R}
\end{equation}
\normalsize

Based on this standardized score, the semantic importance class $L(m)$ is assigned as given in Eq. \ref{e2}.

\small
\begin{equation} \label{e2}
	L(m)=
	\begin{cases}
		\text{high}, & z(m) > 1, \\
		\text{medium}, & -1 \le z(m) \le 1, \\
		\text{low}, & z(m) < -1
	\end{cases}
\end{equation}
\normalsize

\color{black}
The importance class $L(m)$ serves as a control abstraction for downstream modules. Instead of propagating the continuous relevance score $R(m)$ to all layers, the system uses $L(m)$ as a compact cross-layer control signal. In particular, $L(m)$ determines the routing-priority level in the semantic-aware routing module and sets the responsiveness of distortion-control mechanisms under congestion or mobility. This abstraction reduces signaling overhead while preserving the semantic ordering of messages.
\color{black}

The final stage of semantic reasoning determines the appropriate \textit{fidelity level} for message~$m$. Each fidelity level $f \in \mathcal{F}$ corresponds to a different semantic representation granularity in the encoder (e.g., full concept embedding,
reduced feature subset, or minimal semantic signature). Selecting a higher fidelity preserves richer semantic detail but increases representation cost, while lower
fidelity reduces complexity at the expense of expressiveness. To balance these factors, a utility function $\Psi(R(m),f)$ evaluates the benefit of assigning fidelity level~$f$ to message~$m$ based on its relevance score $R(m)$. Messages with higher relevance thus receive richer semantic descriptions, whereas less critical messages can be encoded more compactly.
\color{black}
Using the continuous score $R(m)$ for fidelity selection enables fine-grained adaptation of representation quality, while the discrete level $L(m)$ is reserved for network-layer control decisions. The optimal fidelity decision is formally defined in Eq.~\ref{6}.
\color{black}

\small
\begin{equation}\label{6}
	f^\star(m)
	= \arg\max_{f \in \mathcal{F}} 
	\Psi\big(R(m), f\big)
\end{equation}

\normalsize

The selected fidelity level $f^\star(m)$ is forwarded to the encoder in the Data Plane
\color{black}
and recorded as part of the message’s semantic control state. Together, $f^\star(m)$ and $L(m)$ form a two-dimensional control interface between the semantic reasoning module and the network management modules. Here, $f^\star(m)$ governs representation granularity at the source, while $L(m)$ governs how strongly the network prioritizes and protects the message during forwarding. Accordingly, high-importance messages encoded with high fidelity are steered toward low-distortion paths and granted stronger protection against congestion, whereas lower-importance messages may follow resource-efficient routes. In this way, semantic reasoning directly determines both representation quality and network resource allocation in a coordinated manner.
\color{black}

	\subsection{Semantic-Aware Routing Module}

The semantic-aware routing module selects the forwarding path for each message $m$ by combining its semantic importance with current network conditions. 
It runs in the Knowledge Plane and sends routing decisions to the Data Plane. The module receives the continuous relevance score $R(m)$ and the discrete importance class $L(m)$ from the semantic reasoning stage. These values already capture the effects of the semantic vector $\mathbf{s}(m)$, the task concept set $\mathcal{T}$, the knowledge graph $\mathcal{G}$, and the network state $\mathbf{n}$. Therefore, the routing module does not reprocess semantic content or raw telemetry.

In parallel, the Data Plane provides a set of candidate paths $\mathcal{P}$ for the current source–destination pair. 
\color{black}
To keep routing complexity bounded, $\mathcal{P}$ is not the set of all possible routes. Instead, it is a limited candidate set generated using a constrained multi-path selection method (e.g., a bounded $k$-shortest path strategy based on hop count or delay). This ensures that routing decisions scale with the size of the candidate set rather than the full network size, which is important for real-time operation in large-scale 6G networks.
\color{black}

For each path $p \in \mathcal{P}$, the Data Plane reports three aggregated metrics to the Knowledge Plane:  
(i) predicted end-to-end delay $d(p)$,  
(ii) current path load $\ell(p)$, and  
\color{black}
(iii) predicted semantic distortion $\hat{D}(m,p,f^\star(m))$, which reflects the expected meaning loss when message $m$ is transmitted over path $p$ using the selected fidelity level $f^\star(m)$. Here, $\hat{D}(m,p,f)$ denotes the general distortion model for any fidelity level $f \in \mathcal{F}$, and the routing module evaluates this model at the selected level $f^\star(m)$ provided by the semantic reasoning module. \color{black} Together, these metrics indicate whether a path is fast, stable, and capable of preserving semantic meaning.  
\color{black}
Although the fidelity level $f^\star(m)$ influences the predicted distortion term, it does not directly determine the routing path. The value $f^\star(m)$ is forwarded to the Data Plane encoder to configure the semantic representation granularity, while routing decisions remain driven by semantic relevance and path-level performance metrics. In this way, representation adaptation and path selection remain coordinated but logically separated.
\color{black}

Semantic relevance affects routing sensitivity. Accordingly, when $R(m)$ is high (or $L(m)$ indicates high importance), the routing algorithm becomes more distortion-aware and prioritizes paths with lower $\hat{D}(m,p,f^\star(m))$ and more stable delay. For lower-relevance messages, the algorithm tolerates greater delay or distortion, thereby enabling better load balancing across the network.

Taken together, this module converts semantic importance and network telemetry into a relevance-driven routing decision. The final selected path is denoted by $p^\star(m)$. This decision is pushed to the Data Plane routing agent, which installs the necessary forwarding rules and ensures that message $m$ is transmitted along the chosen semantic-aware route. The routing module computes a semantic-aware path cost in the following two stages.

\begin{algorithm}[h]
\caption{Semantic-Aware Routing Module}
\scriptsize
\label{alg:routing}
\begin{algorithmic}[1]
\Require relevance score $R(m)$, importance class $L(m)$, selected fidelity level $f^\star(m)$, candidate paths $\mathcal{P}$, predicted distortions $\hat{D}(m,p,f^\star(m))$, delays $d(p)$, loads $\ell(p)$

\ForAll{$p \in \mathcal{P}$}
    \State Compute semantic distortion cost $J_1(m,p)$ (Eq.~\ref{7})
    
    \State Compute delay-load cost 
    $J_2(p)$ (Eq.~\ref{8})
    
    \State Compute total routing cost 
    $C(m,p)$ (Eq.~\ref{9})
\EndFor

\State Select optimal path 
$p^\star(m)$ (Eq.~\ref{10})
\State \Return $p^\star(m)$
\end{algorithmic}
\end{algorithm}
    
\subsubsection{Distortion-Weighted Semantic Cost}
In the first stage, the routing module evaluates the expected semantic degradation of each candidate path. The fundamental principle is that tolerance for distortion is inversely proportional to the criticality of the transmitted information. Therefore, the predicted semantic distortion is weighted by the continuous relevance score $R(m)$.
\color{black}
Since distortion depends on both the path and the selected fidelity level, the routing module uses the predicted distortion term $\hat{D}(m,p,f^\star(m))$ provided by the distortion model.
\color{black}
The distortion-aware cost of path $p$ is defined in Eq.~\ref{7}.

\small
\begin{equation}\label{7}
	J_1(m,p) = R(m) \cdot \hat{D}(m,p,f^\star(m))
\end{equation}
\normalsize

A larger value of $J_1(m,p)$ means that path $p$ is more likely to harm the semantic content of message $m$.  
When $R(m)$ is high, even a small distortion leads to a large penalty. When $R(m)$ is low, the same level of distortion is considered more acceptable. This continuous weighting avoids rigid priority boundaries. Highly meaningful messages are routed along semantically reliable paths, whereas less critical messages are permitted to use more resource-efficient routes.

	\subsubsection{Delay-Load Congestion Cost}

In the second stage, the routing module considers end-to-end performance constraints by incorporating latency and congestion along each candidate path. Each path $p$ is described by two telemetry-driven metrics: the predicted delay $d(p)$ and the current load level $\ell(p)$. These metrics are combined using operator-defined weights $\eta_1$ and $\eta_2$, where $\eta_1 + \eta_2 = 1$, as shown in Eq.~\ref{8}.

\small
\begin{equation}\label{8}
	J_2(p) = \eta_1 d(p) + \eta_2 \ell(p)
\end{equation}
\normalsize

This term reflects the classical network performance cost of a path. The weights allow the operator to adapt routing behavior to different service needs. For example, delay can be emphasized in URLLC-like scenarios, while load balancing can be prioritized in high-throughput regimes. By jointly capturing delay and congestion effects, this stage favors paths that are both fast and stable under dynamic 6G conditions.

After computing the distortion-aware cost $J_1(m,p)$ and the performance cost $J_2(p)$, the routing module combines them into a unified semantic-aware routing cost. At this point, the discrete semantic-importance class $L(m)$ determines the relative emphasis on semantic preservation versus network efficiency. This is achieved through a weighting factor $\kappa(L(m))$, which increases with message importance. Here, $\kappa(L(m)) \in [0,1]$ is a class-dependent parameter. High-importance messages use larger $\kappa$ values, making routing more sensitive to semantic distortion. Low-importance messages exhibit smaller $\kappa$ values, allowing the routing decision to place greater emphasis on delay and load efficiency. Accordingly, the final routing cost is defined in Eq.~\ref{9}.

\small
\begin{equation}\label{9}
	C(m,p) = \kappa(L(m)) J_1(m,p)
	+ \big(1-\kappa(L(m))\big) J_2(p)
\end{equation}

\normalsize
Finally, the routing module selects the path with the minimum semantic-aware cost as given in Eq. \ref{10}. The selected path $p^\star(m)$ is translated into forwarding rules and installed in the Data Plane. The message is forwarded along this path under normal conditions. If the distortion control module later detects excessive semantic degradation, a re-routing process can be triggered.

\small
\begin{equation}\label{10}
	p^\star(m) = \arg\min_{p \in \mathcal{P}} C(m,p)
\end{equation}
\normalsize

\color{black}
In large-scale networks, the bounded candidate-path strategy can be combined with hierarchical routing abstractions and event-driven path updates. Candidate paths are recomputed only when significant topology or load changes occur. This prevents routing complexity from scaling with network size and ensures that semantic-aware routing remains feasible under real-time control constraints.
\color{black}

	\subsection{Semantic Distortion Control Module}
The semantic distortion control module ensures that the meaning carried by each semantic message is preserved during transmission. This module operates alongside the semantic-aware routing mechanism and adjusts encoding fidelity or triggers re-routing when semantic degradation becomes unacceptable. All monitoring and decision logic reside in the Knowledge Plane, while corrective actions are executed in the Data Plane. The module consists of two stages: (i) distortion estimation and monitoring, and (ii) relevance-aware corrective control. 

\subsubsection{Distortion Estimation and Monitoring}

Once a path $p^\star(m)$ and fidelity level $f^\star(m)$ are selected for message $m$, the module receives the predicted semantic distortion $\hat{D}(m,p^\star(m),f^\star(m))$ from the Data Plane. After transmission, the observed distortion $D_{\text{obs}}(m)$ is computed at the receiver by comparing the delivered semantic vector $\mathbf{s}'(m)$ with the original representation $\mathbf{s}(m)$, as defined in Eq.~\ref{11}. This comparison reflects the actual semantic degradation introduced by channel impairments, mobility, and congestion effects along the selected path.

\small
\begin{equation}\label{11}
	D_{\text{obs}}(m) =
	1 - \frac{\mathbf{s}(m)\cdot \mathbf{s}'(m)}
	{\|\mathbf{s}(m)\| \|\mathbf{s}'(m)\|}
\end{equation}
\normalsize

The Knowledge Plane then evaluates the mismatch between predicted and observed distortion as given in Eq. \ref{12}. Here, $\Delta(m)$ measures the reliability of the distortion prediction model. A small value indicates that network behavior aligns with expectations, while a large deviation suggests that current channel or mobility dynamics are not well captured by the predictive model.

\small
\begin{equation}\label{12}
	\Delta(m) =
	\left| D_{\text{obs}}(m) -
	\hat{D}(m,p^\star(m),f^\star(m)) \right|
\end{equation}
\normalsize

It is important to emphasize that during the routing phase, predicted distortion $\hat{D}(m,p,f^\star(m))$ is evaluated for each candidate path $p \in \mathcal{P}$. After the optimal path $p^\star(m)$ is selected, the distortion-control mechanism focuses only on the distortion associated with the chosen configuration, namely $\hat{D}(m,p^\star(m),f^\star(m))$. This ensures consistency between routing decisions, fidelity selection, and post-transmission monitoring.
\color{black}

\subsubsection{Relevance-Aware Corrective Control}
After monitoring the distortion behavior, the controller decides whether corrective action is necessary by comparing the distortion gap $\Delta(m)$ with a relevance-dependent tolerance threshold $\delta(m)$. Messages with higher semantic relevance must preserve meaning more strictly. Therefore, the allowable distortion margin decreases as the relevance score increases. To avoid unrealistically strict zero-tolerance decisions and to account for estimation noise in practical systems, a minimum tolerance floor $\delta_{\min}$ is introduced. The relevance-aware tolerance is therefore defined as

\begin{equation}\label{13}
\delta(m) = \delta_{\min} + (\delta_0 - \delta_{\min})(1 - R(m)).
\end{equation}

Here, $\delta_0$ denotes the maximum distortion margin tolerated for very low-relevance messages. It is determined during a short calibration phase under nominal network conditions by measuring typical semantic distortion, thereby capturing the deployment’s baseline semantic tolerance. The parameter $\delta_{\min}$ defines the minimum admissible margin for highly relevant messages, preventing control oscillations due to noise or prediction errors while still enforcing strict semantic preservation. If $\Delta(m) \le \delta(m)$, the observed distortion is considered acceptable and no corrective action is required, meaning the current routing path $p^\star(m)$ and encoding fidelity $f^\star(m)$ adequately preserve the task-relevant meaning. Otherwise, the controller first attempts to maintain semantic fidelity before resorting to more disruptive actions such as rerouting.

\begin{equation}\label{14}
\color{black}
f_{\text{temp}}(m) =
f^\star(m) + \lambda \,\Delta(m)
\color{black}
\end{equation}

To address excess distortion, the control logic first attempts correction at the representation level, since adjusting fidelity is less disruptive than changing the end-to-end path. The controller computes a tentative fidelity update proportional to the distortion gap (Eq.~\ref{14}), where $\lambda \in (0,1)$ is a small adaptation gain controlling the reaction to unexpected semantic degradation. Because the encoder supports only discrete semantic configurations $\mathcal{F} = \{f_{\text{low}}, f_{\text{mid}}, f_{\text{high}}\}$, the temporary value $f_{\text{temp}}(m)$ is projected onto the nearest admissible level to obtain $f_{\text{new}}(m)$. If this level provides stronger protection than the current fidelity $f^\star(m)$, the Knowledge Plane instructs the Data Plane encoder to adopt $f_{\text{new}}(m)$ for subsequent transmissions while keeping the existing route. If the maximum fidelity is already in use and $\Delta(m)$ still exceeds the allowable margin, the controller determines that the current path cannot reliably preserve semantic meaning and triggers re-routing, prompting the routing module to recompute $p^\star(m)$ under updated network conditions. This two-step strategy limits path oscillations while maintaining semantic reliability in dynamic 6G environments.

		\begin{algorithm}[h]
\caption{Semantic Distortion Control Module}
\scriptsize
\label{alg:distortion}
\begin{algorithmic}[1]
\Require $R(m)$, $f^\star(m)$, $\hat{D}(m,p^\star(m),f^\star(m))$, $D_{\text{obs}}(m)$ 

\State Compute $\Delta(m)$ using Eq.~\ref{12}
\State Compute $\delta(m)$ using Eq.~\ref{13}

\If{$\Delta(m) \le \delta(m)$}
    \State \Return $f^\star(m)$ \hfill \textit{(no correction needed)}
\Else
    \State Compute tentative fidelity $f_{\text{temp}}(m)$ using Eq.~\ref{14}
    \State Project $f_{\text{temp}}(m)$ onto nearest admissible level $f_{\text{new}}(m) \in \mathcal{F}$

    \If{$f_{\text{new}}(m) > f^\star(m)$}
        \State Apply $f_{\text{new}}(m)$ in the Data Plane encoder
        \State \Return $f_{\text{new}}(m)$
    \Else
        \State Trigger re-routing request for message $m$
        \State \Return \texttt{ReRoute}$(m)$
    \EndIf
\EndIf
\end{algorithmic}
\end{algorithm}

The three modules operate as a unified control loop enabling semantic-aware communication in the proposed 6G management architecture. The semantic reasoning module first evaluates message importance by combining task relevance, contextual relations, and network urgency into a relevance score. Together with the resulting importance class and fidelity level, this score serves as the main input to the semantic-aware routing module, which selects paths by jointly considering predicted semantic distortion, end-to-end delay, and traffic load. After transmission, the semantic distortion control module compares predicted and observed distortion (Algorithm 3) and, when necessary, increases encoding fidelity or triggers re-routing to preserve semantic meaning under dynamic conditions.

Through this interaction, the modules form an integrated decision pipeline: reasoning determines message importance, routing allocates resources accordingly, and distortion control continuously monitors and corrects semantic quality. This enables the network to adapt representation fidelity and forwarding decisions in real time while maintaining efficient resource use and task-relevant meaning across heterogeneous and time-varying 6G environments. From a control perspective, the framework behaves as a bounded feedback-driven adaptation loop rather than a strict optimization controller. Reasoning updates, routing changes, and fidelity adjustments occur only when predicted or observed distortion exceeds predefined thresholds, limiting reactions to short-term fluctuations. Relevance-weighted updates, temporally smoothed telemetry, and discrete control intervals further reduce oscillations and route flapping under mobility and load dynamics. Corrective actions are therefore triggered only for meaningful semantic deviations, keeping control overhead low while preserving stability and responsiveness. Although analytical optimality is not claimed, the framework encourages convergence toward a stable operating region, reflected in bounded distortion trends, fewer re-routing events, and faster stabilization observed in the evaluation.

\section{Performance Evaluation}
\subsection{Simulation Environment}
To evaluate the proposed semantic communication framework, we construct a simulation environment that integrates the ns-3 network simulator with a Python-based Knowledge Plane. This setup reproduces the complete semantic control loop defined in the system model. All modules operate online using telemetry collected from the Data Plane, enabling an end-to-end assessment of how semantic decisions affect communication performance under multi-hop 6G network conditions. The architectural framework of the simulation environment is organized according to the following specifications:
\begin{table}[h]
\centering
\caption{Network, Traffic, and Topology Parameters}
\label{tab:net_params}
\renewcommand{\arraystretch}{1.1}
\scriptsize
\begin{tabular}{p{2.7cm} || p{4.6cm}}
\hline
\textbf{Parameter} & \textbf{Value / Description} \\ \hline
\hline
Simulator & ns-3.38 \\ \hline
Topology & Heterogeneous multi-hop 6G access network \\ \hline
Area size & $1\,\text{km}^2$ \\ \hline
Node types & Macro BSs, small cells, relays, access points \\ \hline
UE mobility model & Random Waypoint, 1--15 m/s \\ \hline
Link model & SINR-driven stochastic capacity model \\ \hline
Capacity range & 80--900 Mbps (emergent from SINR) \\ \hline
Path loss model & 3GPP UMa-inspired large-scale model \\ \hline
Shadowing & Log-normal (stochastic) \\ \hline
Interference & Concurrent transmission-based (time-varying) \\ \hline
Queueing model & Packet-level FIFO abstraction per node \\ \hline
Propagation delay & 1--3 ms per hop (stochastic) \\ \hline
Background traffic & Bursty ON/OFF CBR, 20--60 ms periods \\ \hline
Semantic flow arrivals & Poisson, $\lambda_m = 15$ messages/s \\ \hline
Semantic packet size & 512--1024 bytes (post-encoding) \\ \hline
Telemetry interval & 0.2 s \\ \hline
Control interval & 0.2 s (aligned with telemetry) \\ \hline
Simulation duration & 180 s (20 s warm-up + 160 s evaluation) \\ \hline
Repetitions & 10 independent runs per scenario \\ \hline
\end{tabular}
\end{table}

\subsubsection{Network Topology and Physical Layout}
We simulate a heterogeneous multi-hop 6G access network in ns-3, where macro base stations, small cells, relays, and access points are deployed in a spatially irregular manner over a $1\,\text{km}^2$ area. User devices follow a Random Waypoint mobility model, creating time-varying link conditions and routing dynamics. Wireless links are modeled using an SINR-driven stochastic capacity abstraction that captures path loss, shadowing, and interference effects. Packet-level queueing is included to reflect congestion behavior while keeping the focus on semantic control mechanisms. The complete set of network, traffic, and topology parameters is summarized in Table~\ref{tab:net_params}.

\subsubsection {Semantic Message Generation}  
Semantic messages arrive at each source according to a Poisson process with rate $\lambda_m = 15$ messages/s. Each message is represented by a 128-dimensional semantic vector produced by a lightweight pre-trained transformer-based embedding model executed externally in Python. The model is used solely for feature extraction and no additional training is performed within the simulator. During simulation, task-specific message templates are encoded into semantic vectors and provided to ns-3 as fixed representations. Messages are associated with task categories that define the corresponding concept set $\mathcal{T}$ used by the semantic relevance module. This setup allows evaluation of the proposed semantic control mechanisms while keeping the embedding process constant. To avoid redundancy, the main semantic and control parameters are summarized in Table~\ref{tab:sem_params}.

\begin{table}[h]
\centering
\caption{Semantic, Routing, and Control Parameters}
\label{tab:sem_params}
\renewcommand{\arraystretch}{1.1}
\scriptsize
\begin{tabular}{p{3.7cm} || p{3.8cm}}
\hline
\textbf{Parameter} & \textbf{Value / Description} \\ \hline\hline
Semantic vector dimension & 128 \\ \hline
Semantic fidelity set $\mathcal{F}$ & $\{f_{\text{low}}, f_{\text{mid}}, f_{\text{high}}\}$ \\ \hline
Knowledge graph size & 200--500 concepts (offline constructed) \\ \hline
Local reasoning scope & Bounded neighborhood $\mathcal{N}(m)$ \\ \hline
Encoding latency & 1--5 ms (edge-class inference) \\ \hline
Distortion metric & Cosine distance (Eq.~\ref{11}) \\ \hline
Relevance fusion weights $(\alpha,\beta,\gamma)$ & (0.4, 0.3, 0.3) in Eq.~\ref{4} \\ \hline
Routing delay/load weights $(\eta_1,\eta_2)$ & (0.5, 0.5) in Eq.~\ref{8} \\ \hline
Importance–routing weight $\kappa(L)$ & \{0.7 (high), 0.5 (mid), 0.3 (low)\} \\ \hline
Base distortion tolerance $\delta_0$ & 0.05 \\ \hline
Fidelity adaptation gain $\lambda$ & 0.1, bounded in $(0,1)$ \\ \hline
Re-routing trigger & $\Delta(m) > \delta(m)$ after fidelity projection \\ \hline
Concurrent semantic flows & 20--40 \\ \hline
Background load level & 30--70\% of instantaneous link capacity \\ \hline
Control interval & 0.2 s (closed-loop cycle) \\ \hline
\end{tabular}
\end{table}

\subsubsection{Telemetry and Distortion Monitoring}
During forwarding, nodes record per-hop metrics such as delay, queue state, load, and link quality, which are periodically exported to the Knowledge Plane through ns-3 trace callbacks. Upon reception, the observed semantic distortion $D_{\text{obs}}(m)$ is computed as the cosine distance between the original semantic vector $\mathbf{s}(m)$ and the reconstructed vector $\mathbf{s}'(m)$ reflecting fidelity selection and network effects. The Knowledge Plane compares predicted and observed distortion to trigger fidelity adaptation or rerouting when necessary. The relevant monitoring and control parameters are summarized in Table~\ref{tab:sem_params}.

\subsubsection {Knowledge Plane Integration}  
The Knowledge Plane is implemented as a standalone Python process interfaced with ns-3 through control and telemetry APIs. It receives periodic telemetry updates, semantic vectors, and post-delivery distortion measurements from the Data Plane. Upon message generation, the semantic reasoning module computes the relevance score $R(m)$, assigns an importance class $L(m)$, and determines the target fidelity level $f^\star(m)$. Based on these outputs and current network state information, the semantic-aware routing module evaluates candidate paths using the proposed semantic-aware cost functions and selects the optimal forwarding path $p^\star(m)$. Routing decisions are enforced in ns-3 via dynamic route table updates. After message delivery, the semantic distortion control module compares predicted distortion with the observed distortion value. If a deviation beyond the admissible threshold is detected, the Knowledge Plane adjusts the fidelity configuration or triggers a re-routing request for subsequent transmissions. All computations are performed online at each telemetry interval, forming a closed-loop semantic control structure aligned with KDN.

\subsubsection {Traffic Model and Load Dynamics}  
The network carries both semantic flows and background traffic to emulate realistic load fluctuations. Background traffic is generated using a bursty CBR-based ON/OFF model, where ON and OFF durations are uniformly distributed between 20--60~ms. During ON periods, nodes transmit fixed-rate traffic streams, creating transient congestion and queue buildup. This time-varying load environment stresses the adaptive behavior of semantic-aware routing and distortion control mechanisms under dynamic congestion conditions.

\subsubsection {PHY/MAC Abstraction and Modeling Scope}
To evaluate network-level semantic decision-making without excessive physical-layer complexity, the PHY and MAC layers are abstracted using an SINR-driven link model. Instantaneous link capacity is computed using the Shannon-based approximation
$R_{ij}(t) = B \log_2(1 + \text{SINR}_{ij}(t))$, where $\text{SINR}_{ij}(t)$ reflects path loss, shadowing, and stochastic interference from concurrent transmissions. These variations generate realistic fluctuations in link capacity that propagate to queueing delay and congestion at the network layer. This abstraction preserves the interaction between wireless dynamics and routing behavior while maintaining computational tractability. The goal of the study is not PHY optimization, but to analyze how semantic-aware routing and closed-loop distortion control react to changing network conditions.

\subsubsection{Knowledge Graph Construction and Scalability Assumptions}
The knowledge graph $\mathcal{G}$ is constructed offline from operator ontologies and application concept models. It follows a lightweight schema where nodes represent task-related concepts and edges encode semantic relations such as hierarchy, dependency, similarity, or contextual association. Concept embeddings are aligned with the message representation space, and the semantic affinity $w_{m,v}$ between a message $m$ and a neighboring concept $v$ is computed using cosine similarity between $\mathbf{s}(m)$ and the embedding of $v$, producing a normalized relevance weight in $[0,1]$. The graph is validated offline and remains static during simulation. To ensure scalability, semantic reasoning is restricted to a bounded local neighborhood $\mathcal{N}(m)$ rather than global traversal, making computation dependent on local connectivity instead of total graph size. In our experiments the graph contains a few hundred sparsely connected concepts, which is sufficient to represent task semantics while keeping reasoning efficient. Routing scalability is further maintained by evaluating only a bounded candidate path set $\mathcal{P}$ from the Data Plane, so decision complexity depends on $|\mathcal{P}|$ rather than the overall network size.

\subsubsection{Simulation Duration and Randomization}
Each simulation lasts 180~s and is divided into three phases representing progressively challenging conditions. P1 (baseline) operates under moderate load and stable mobility; P2 (stress) increases traffic intensity and mobility, causing congestion bursts and link-quality variations; and P3 (highly dynamic) combines sustained load with persistent mobility, producing rapidly changing paths and higher semantic distortion risk. Each scenario is repeated 10 times with independent random seeds that randomize node placement perturbations, mobility trajectories, fading realizations, and traffic patterns. The first 20~s are used as warm-up, leaving 160~s for evaluation. Across runs, the semantic control loop stabilizes within 3--6 control intervals after disturbances. Reported results are means with 95\% confidence intervals to ensure statistical robustness.

\subsubsection{Deployment Considerations and Operational Robustness}
From a deployment perspective, the knowledge graph is constructed offline using operator service specifications, application metadata, and historical traffic logs, with concept relations derived from ontologies or embedding similarity and task concept sets defined during service provisioning. Incremental updates can be applied during low-load periods to incorporate new services without full reconstruction. During operation, the Knowledge Plane runs at a 0.2~s control interval, much slower than packet forwarding. If it becomes temporarily unavailable, the Data Plane continues using the latest routing paths and fidelity settings, since decisions occur per control cycle rather than per packet. Delayed telemetry preserves the last stable configuration, while prolonged outages trigger graceful degradation to load-aware routing with fixed fidelity, maintaining stable packet delivery without semantic optimization.

\begin{table*}[h]
\centering
\caption{Baseline Schemes and Isolated Design Dimensions}
\label{tab:baseline_summary}
\renewcommand{\arraystretch}{1}
\scriptsize
\begin{tabular}{l||cccc}
\hline
\textbf{Scheme} 
& \textbf{Semantic Relevance} 
& \textbf{Distortion Prediction} 
& \textbf{Load Awareness} 
& \textbf{Closed-Loop Control} \\
\hline
\hline
SP 
& \xmark 
& \xmark 
& \xmark 
& \xmark \\
\hline
LBR 
& \xmark 
& \xmark 
& \cmark 
& \xmark \\
\hline
DMR 
& \xmark 
& \cmark 
& \xmark 
& \xmark \\
\hline
Proposed Framework
& \cmark 
& \cmark 
& \cmark 
& \cmark \\
\hline
\end{tabular}
\end{table*}

    \begin{figure}[h]
		\centering
		\includegraphics[width=\linewidth]{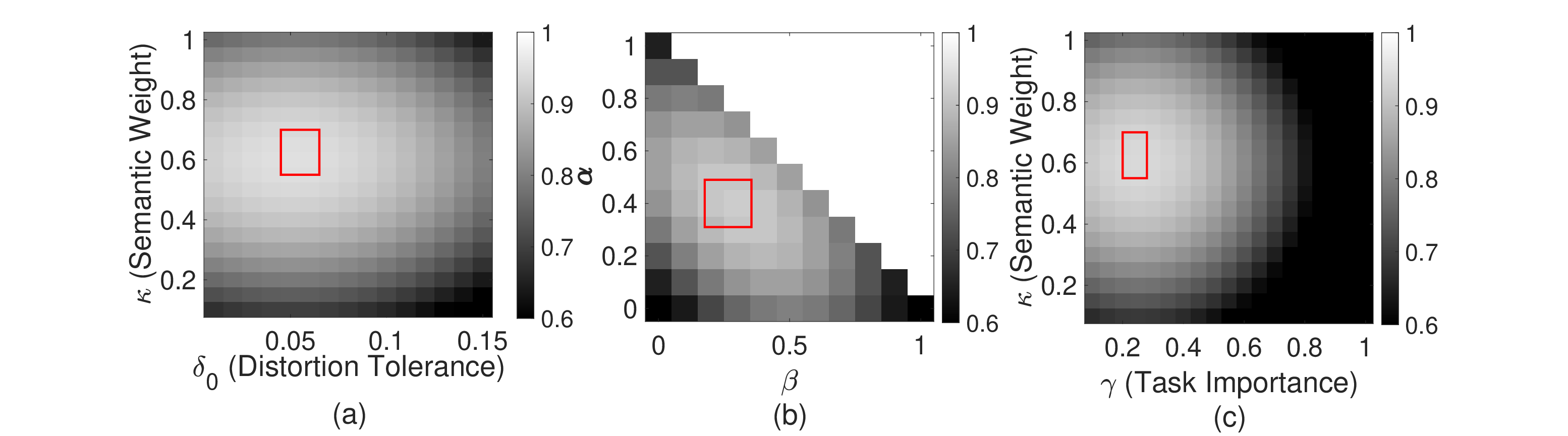}	
		\caption{Parameter sensitivity and selection rationale}
		\label{f8}
	\end{figure}
    
\subsection{Parameter Sensitivity and Selection Rationale}
To avoid arbitrary tuning, we conduct a sensitivity analysis over the main control parameters of the semantic-control loop. Fig.~\ref{f8} illustrates the impact on the Semantic Delivery Success Rate (SDSR) of (a) the routing weight $\kappa$ and base distortion tolerance $\delta_0$, (b) the relevance weights $(\alpha,\beta,\gamma)$ with $\alpha+\beta+\gamma=1$, and (c) the interaction between $\kappa$ and the task-importance factor $\gamma$.

Fig.~\ref{f8}a shows that SDSR reaches its highest values around $\kappa \approx 0.6$ and $\delta_0 \approx 0.05$, forming a broad plateau rather than a sharp optimum. Very small $\delta_0$ makes the controller overly reactive, whereas large $\delta_0$ delays corrective actions and increases accumulated semantic distortion. The selected operating point lies near the center of this stable high-performance region. In Fig.~\ref{f8}b, SDSR is evaluated over the simplex defined by $\alpha$ and $\beta$, with $\gamma = 1-\alpha-\beta$. The highest values occur near $(\alpha,\beta)\approx(0.4,0.3)$, implying $\gamma\approx0.3$. The results indicate that moderate differentiation among relevance components performs better than either extreme or uniform weighting. Fig.~\ref{f8}c examines the interaction between $\kappa$ and $\gamma$. The best region appears near $(\kappa,\gamma)\approx(0.6,0.25)$. SDSR is more sensitive to $\kappa$, indicating that distortion-aware routing dominates performance while task importance acts as a secondary modulation. The smooth transitions across the surface suggest stable control behavior across the explored parameter space.

Additional parameters are selected for stability and implementation practicality. Delay–load weights are fixed at $(\eta_1,\eta_2)=(0.5,0.5)$ to balance latency and throughput. The fidelity adaptation gain is set to $\lambda=0.1$ to ensure gradual corrections and avoid oscillations. Semantic importance is derived using a $\pm1\sigma$ threshold over standardized relevance scores, while the encoder employs three discrete fidelity levels to limit signaling overhead. The urgency mapping is normalized to $U(m)\in[0,1]$ for consistent scaling. Together with bounded-neighborhood reasoning in the knowledge graph and restricted candidate-path evaluation, these settings keep the total control latency within the 0.2~s telemetry interval, allowing online operation of the semantic management loop.

\subsection{Baseline Methodologies}

To evaluate the impact of semantic reasoning, semantic-aware routing, and closed-loop distortion control, the proposed framework is compared with three representative routing baselines reflecting common networking paradigms. Since no prior work provides a directly comparable multi-hop semantic routing architecture, these baselines isolate different design principles, enabling controlled comparison of semantic and network-level behavior as summarized in Table~\ref{tab:baseline_summary}.

\begin{itemize}

\item {Shortest-Path Routing (SP):}
Selects the minimum-hop path between source and destination. It ignores semantic relevance, predicted distortion, and load conditions, serving as a conventional topology-driven lower-bound benchmark.

\item {Load-Based Routing (LBR):}
Chooses the path with the lowest instantaneous load. Although congestion-aware, it does not incorporate semantic relevance or distortion prediction and reflects classical traffic-engineering strategies.

\item {Distortion-Minimizing Routing (DMR):}
Selects the path with minimum predicted semantic distortion. While prioritizing semantic fidelity, it ignores delay, congestion, and relevance-based prioritization, representing a partial semantic approach without closed-loop adaptation.

\end{itemize}

\subsection{Evaluation Metrics}
To evaluate the proposed framework, we consider a set of complementary metrics capturing semantic fidelity, routing dynamics, and network-level efficiency. As summarized in Table~\ref{tab:metric_summary}, each metric is aligned with a specific design dimension of the semantic-control architecture. Together, these metrics assess how effectively semantic meaning is preserved while maintaining stable routing behavior and efficient network utilization.

\begin{itemize}

\item {End-to-End Semantic Distortion:}  
This metric quantifies semantic deviation introduced during transmission. The observed distortion is computed as the cosine distance between the original embedding $\mathbf{s}(m)$ and the reconstructed embedding $\mathbf{s}'(m)$ at the receiver $D_{\text{obs}}(m) = 1 - \frac{\mathbf{s}(m)\cdot \mathbf{s}'(m)}{\|\mathbf{s}(m)\| \|\mathbf{s}'(m)\|}$. Here, lower values indicate better preservation of task-relevant semantic content.

\item {Semantic Delivery Success Rate (SDSR):}  
A semantic message is considered successfully delivered if its observed distortion satisfies the relevance-dependent tolerance $D_{\text{obs}}(m) \le \delta(m), \quad
\delta(m) = \delta_0 \big(1 - R(m)\big)$, where $R(m)\in[0,1]$ denotes semantic relevance and $\delta_0$ is the baseline distortion tolerance. This success criterion is applied identically across all evaluated schemes to ensure consistent comparison. SDSR is defined as the fraction of successfully delivered semantic messages.

\item {End-to-End Delay:}  
Average time required for a semantic message to reach its destination, including queueing, propagation, transmission, and processing delays. This metric reflects the latency implications of semantic-aware routing and control adaptation.
 \begin{table}[h]
\centering
\caption{Evaluation Metrics and Design Dimensions}
\label{tab:metric_summary}
\renewcommand{\arraystretch}{1}
\tiny
\begin{tabular}{p{2.3cm} || p{1.4cm} p{1.4cm} p{1.4cm}}
\hline
\textbf{Metric} 
& \textbf{Semantic Fidelity} 
& \textbf{Routing Dynamics} 
& \textbf{Network Efficiency} \\
\hline
\hline
End-to-End Semantic Distortion 
& \cmark 
& \xmark 
& \xmark \\
\hline
Semantic Delivery Success Rate 
& \cmark 
& \cmark 
& \cmark \\
\hline
End-to-End Delay 
& \xmark 
& \cmark 
& \cmark \\
\hline
Path Stability 
& \xmark 
& \cmark 
& \xmark \\
\hline
Network Throughput 
& \xmark 
& \xmark 
& \cmark \\
\hline
\end{tabular}
\end{table}

\item {Path Stability:}  
For each semantic flow $f$, the number of control-triggered routing changes $N_{\text{reroute}}(f)$ is recorded. The reported value corresponds to the average re-routing frequency per flow. Lower values indicate more stable routing behavior and reduced control-induced oscillation.

\item {Network Throughput:}  
Aggregate successfully delivered data rate across both semantic and background traffic. This metric evaluates how semantic-aware decisions influence overall link utilization and congestion dynamics.

\end{itemize}

These metrics provide a balanced assessment of semantic-level correctness and classical network performance, ensuring that improvements in meaning preservation do not come at the expense of stability or efficiency.

    \begin{figure*}[h] 
    \centering \includegraphics[width=0.7\linewidth]{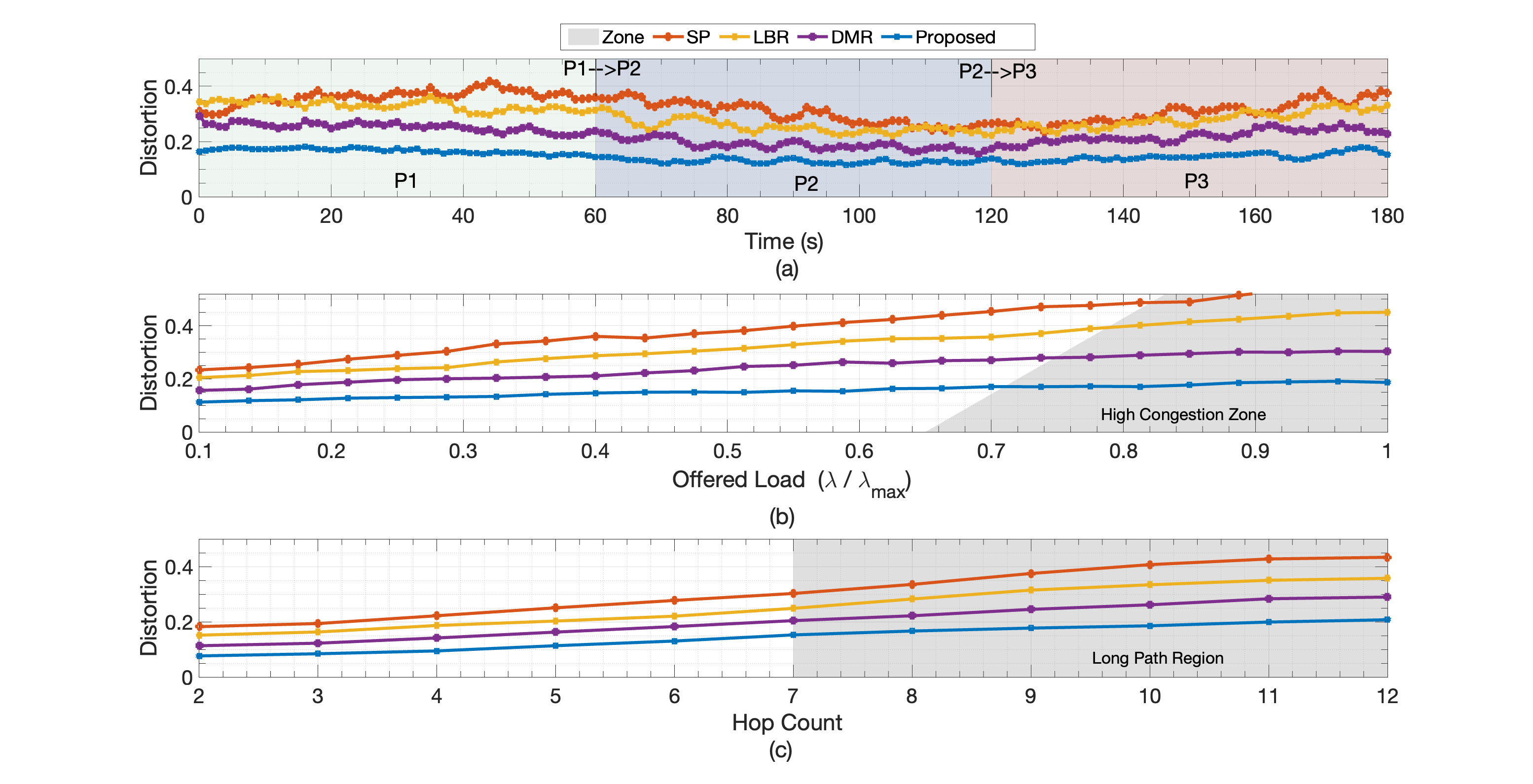} 
    \caption{Evaluations for end-to-end semantic distortion according to a) Time, b) Offered Load, and c) Hop Count} 
    \label{fig4} 
    \end{figure*}
    
\subsection{Simulation Results}
\subsubsection{End-to-End Semantic Distortion}
Fig.~\ref{fig4}a shows the temporal evolution of end-to-end semantic distortion across phases P1–P3. SP exhibits the largest variability, particularly during the P2 stress phase, since topology-only routing cannot react to congestion or mobility-driven instability. LBR reduces fluctuations during P1 but distortion increases in P2 as mobility alters traffic distribution and semantically important flows are not prioritized. DMR behaves more smoothly under stable conditions due to distortion-aware path selection, yet distortion rises near phase transitions where previously favorable paths become unstable. In contrast, the proposed framework maintains the lowest and most stable distortion throughout all phases by jointly considering predicted distortion, semantic relevance, and current network state in routing and fidelity control. Fig.~\ref{fig4}b presents distortion versus offered load. As congestion increases, SP rises sharply due to queue buildup on minimum-hop paths. LBR performs better at moderate loads but becomes unstable under heavy congestion, while DMR increases more gradually yet still reflects congestion sensitivity. The proposed framework shows the flattest growth and lowest overall distortion, as adaptive routing and fidelity control prevent congestion from translating into sustained semantic degradation. Fig.~\ref{fig4}c illustrates distortion with respect to hop count. Distortion increases for all schemes due to cumulative multi-hop effects, but the growth differs. SP increases most rapidly beyond mid-range paths, LBR mitigates distortion on shorter routes but degrades on longer ones, and DMR remains stable up to moderate hops before increasing without congestion awareness. The proposed approach exhibits the slowest and most controlled growth, remaining consistently below the baselines even at larger hop counts. Overall, the framework reduces both average distortion and variability across time, load, and path length.

	\begin{figure*}[h]
		\centering
		\includegraphics[width=0.7\linewidth]{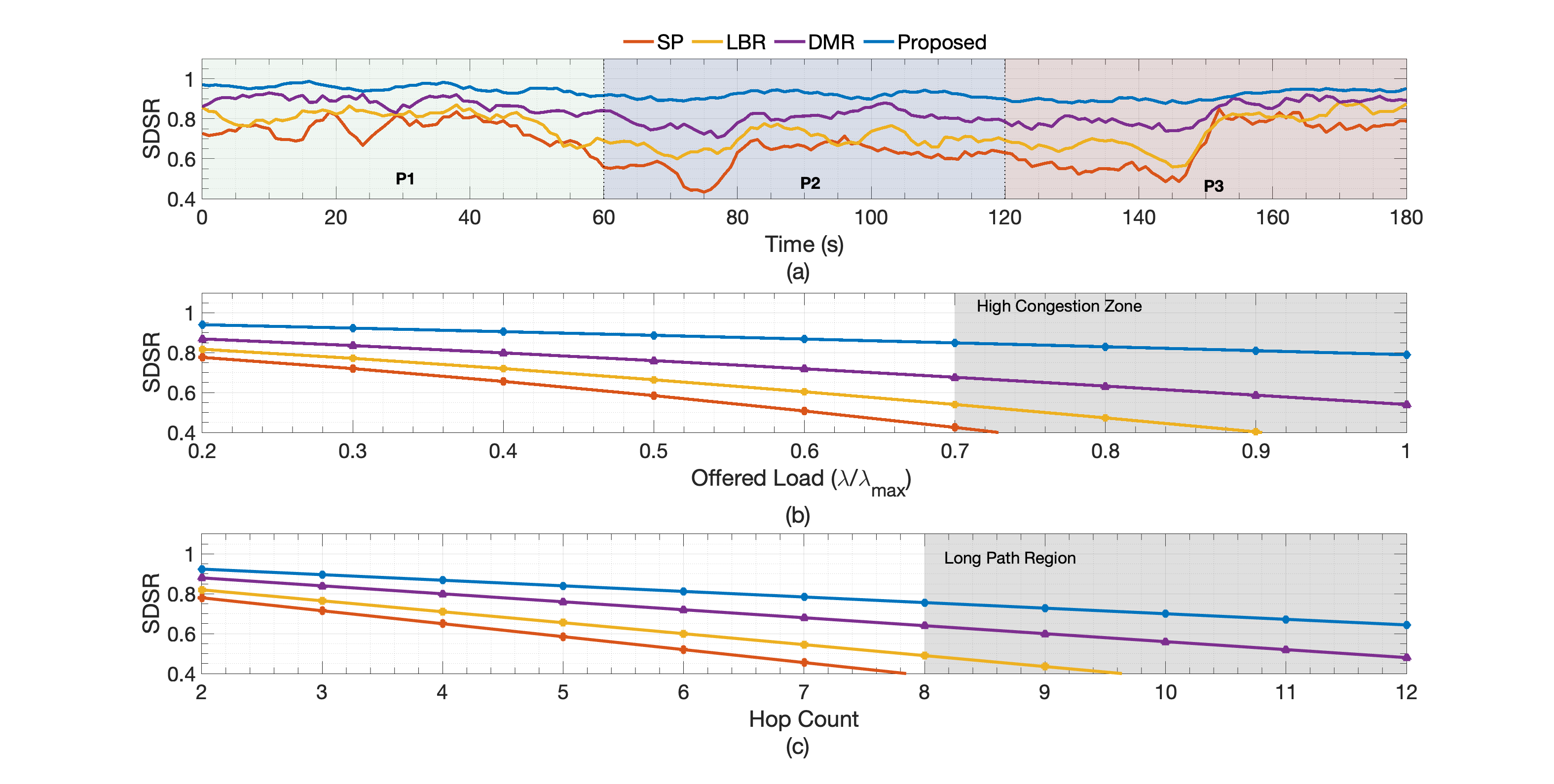}	
		\caption{Evaluations for semantic delivery success rate according to a) Time, b) Offered Load, and Hop count}
		\label{f2}
	\end{figure*}  
        \begin{figure*}[h]
		\centering
		\includegraphics[width=0.7\textwidth]{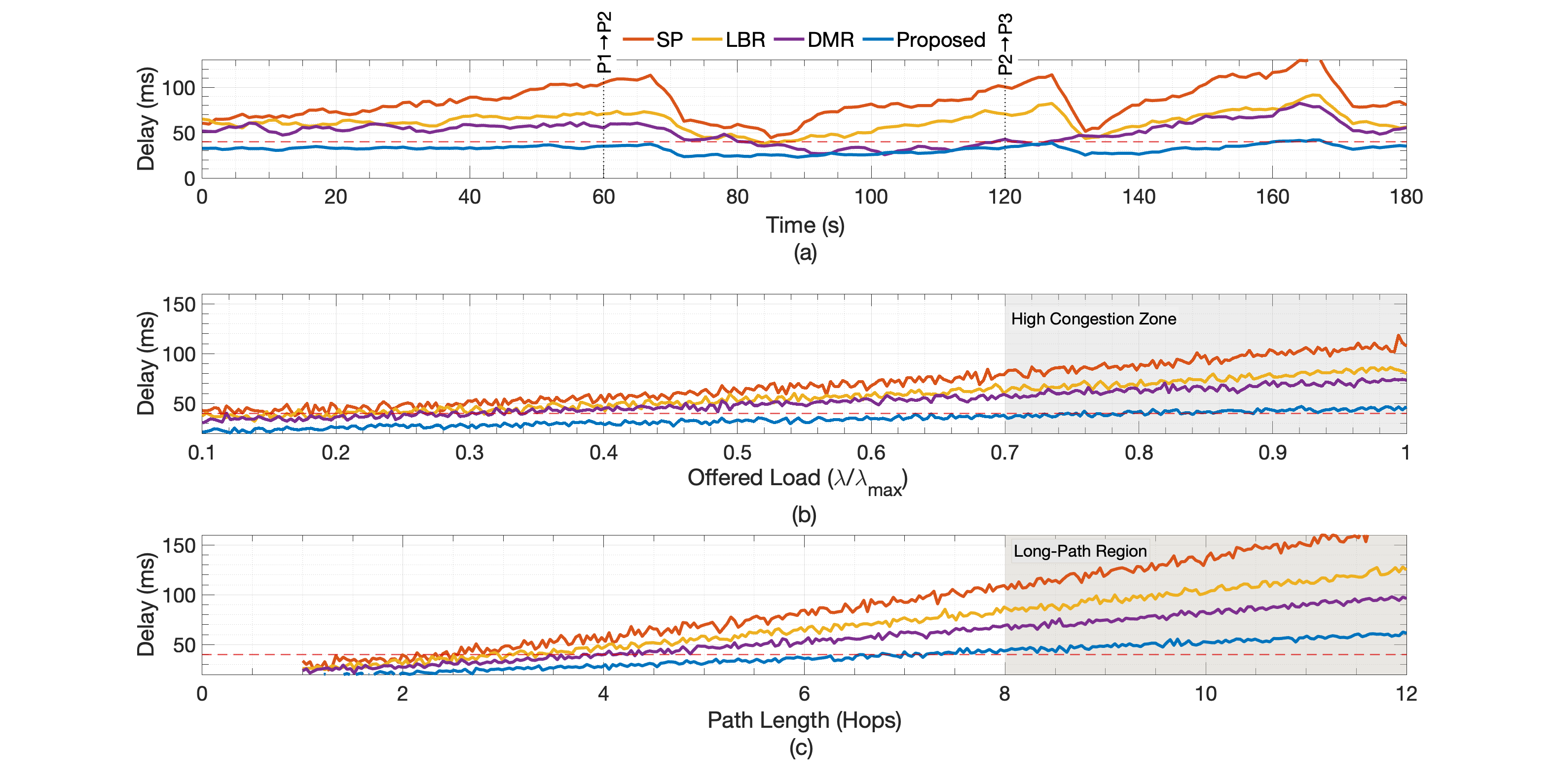}	
		\caption{Evaluations for end-to-end delay}
		\label{f4}
	\end{figure*}

        	\begin{figure*}[h]
		\centering
		\includegraphics[width=0.7\textwidth]{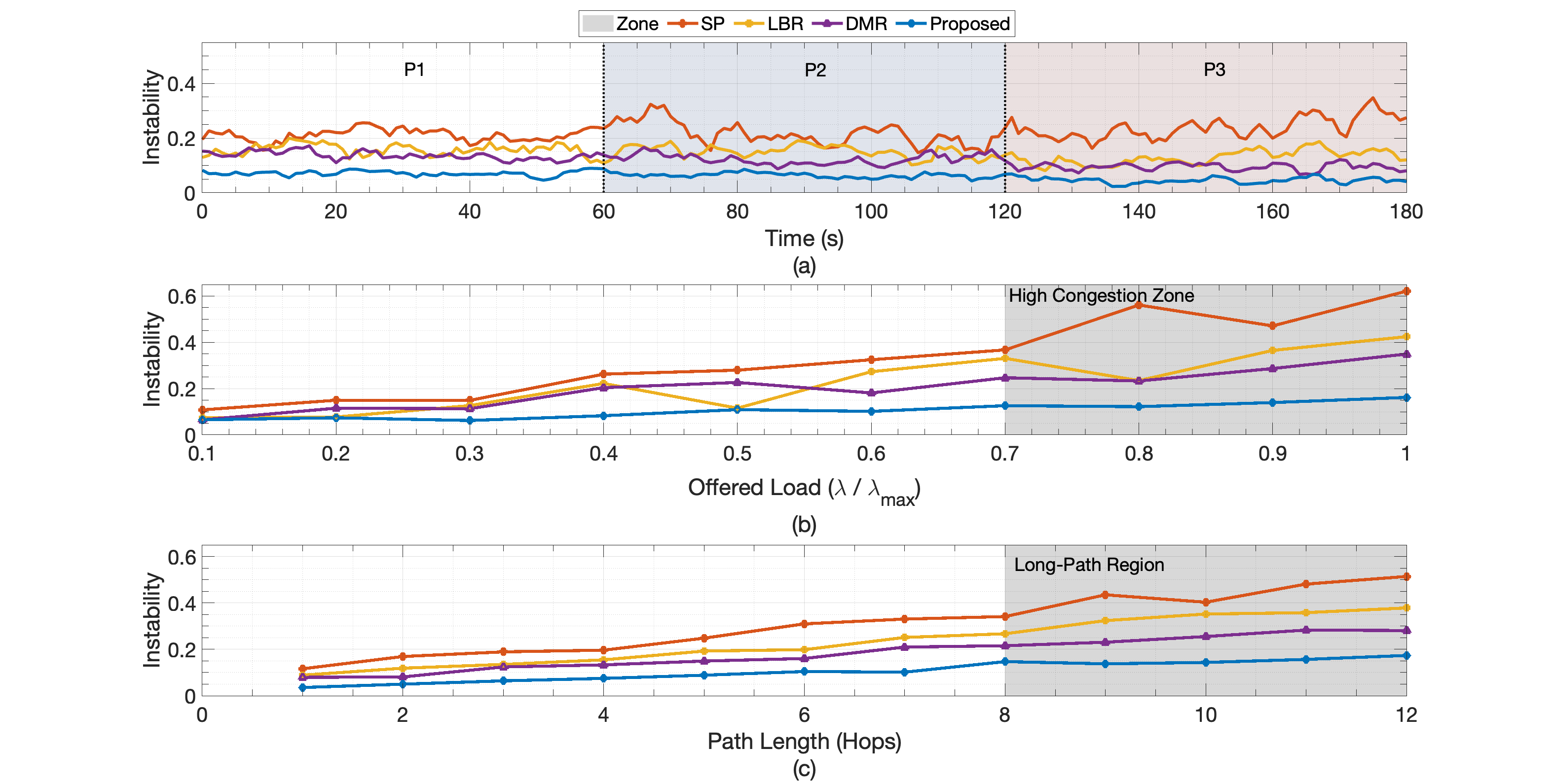}	
		\caption{Evaluations for path stability}
		\label{f5}
	\end{figure*}

\subsubsection{Semantic Delivery Success Rate (SDSR)}
Fig.~\ref{f2}a--\ref{f2}c show the evolution of SDSR across time, offered load, and hop count. In the steady phase (P1) of Fig.~\ref{f2}a, all schemes maintain relatively stable SDSR, although SP and LBR already exhibit minor oscillations, indicating that topology- or load-driven decisions react only after semantic degradation occurs. During the stress phase (P2), mobility and congestion variability cause a visible drop for SP and a moderate decline for LBR. DMR remains more stable by favoring low-distortion paths, yet its SDSR still fluctuates when those paths become congested or unstable. In contrast, the proposed framework maintains consistently higher SDSR across P1–P3 by combining relevance-aware routing with continuous distortion monitoring and proactive adjustments of routing or encoding fidelity. Fig.~\ref{f2}b presents SDSR as a function of offered load. Increasing congestion reduces SDSR for all schemes, with the sharpest decline observed for SP due to queue buildup along minimum-hop paths. LBR mitigates this effect at moderate loads but deteriorates once congestion spreads across the network. DMR degrades more gradually but remains vulnerable when distortion-favorable paths become bottlenecks. The proposed framework exhibits the most stable behavior and the slowest decline, as joint adaptation of routing and fidelity absorbs part of the congestion impact and prevents persistent semantic violations. SDSR versus hop count is illustrated in Fig.~\ref{f2}c. Reliability decreases with path length for all schemes due to cumulative distortion. SP and LBR degrade steadily without modeling semantic drift, while DMR performs better at moderate hop counts before declining under mobility and queue dynamics. The proposed approach shows the slowest reduction in SDSR, since relevance-aware routing avoids assigning long or unstable paths to critical messages and adaptive fidelity compensates for accumulated distortion.

These results indicate that topology-driven routing, load balancing, or distortion-only optimization alone cannot sustain reliable semantic delivery in dynamic 6G environments. High SDSR requires coordinated interaction between semantic reasoning, routing decisions, and closed-loop distortion control, highlighting the systemic benefit of the proposed framework.

\subsubsection{End-to-End Delay}

Delay behavior further illustrates how routing strategies react to mobility and congestion dynamics. Fig.~\ref{f4}a shows the temporal evolution of delay across phases P1–P3. During the initial steady phase (P1), all schemes operate in a relatively stable regime, although SP exhibits a slightly higher baseline delay due to traffic concentration on minimum-hop links. With the transition to the stress phase (P2), mobility and load increase queue buildup. SP experiences the largest delay growth, while LBR partially mitigates congestion through load redistribution but becomes less effective as congestion spreads. DMR remains comparatively smoother by favoring low-distortion paths, yet delay still increases when those paths become overloaded. The proposed framework shows the most stable trajectory, as predictive distortion indicators and network-state awareness limit abrupt rerouting and excessive queue buildup. In the highly dynamic phase (P3), delay fluctuations intensify for SP and LBR, whereas the proposed method maintains comparatively bounded variations and a smoother overall trend. Fig.~\ref{f4}b presents delay as a function of offered load. Delay increases nonlinearly for all schemes, consistent with queueing effects. SP exhibits the steepest growth as central links saturate early. LBR performs better at moderate load levels but deteriorates once congestion becomes widespread. DMR shows a more gradual increase but still rises in the heavy-load regime since distortion minimization alone cannot prevent queue accumulation. The proposed method maintains the flattest curve, as joint consideration of semantic relevance, predicted distortion, and congestion indicators moderates queue growth and balances delay with semantic fidelity. The delay versus hop count is illustrated in Fig.~\ref{f4}c . Delay increases for all schemes due to cumulative multi-hop effects, but the growth patterns differ. SP grows most rapidly beyond mid-range paths, LBR mitigates delay on shorter routes but loses effectiveness on longer ones, and DMR shows moderate delay before increasing in the long-path region. The proposed framework exhibits the slowest growth, since relevance-aware routing avoids assigning long or unstable paths to critical traffic while adaptive fidelity helps limit queue buildup. Overall, the results indicate that stable latency in semantic communication requires coordinated interaction between semantic reasoning and network-state awareness rather than topology-, load-, or distortion-only optimization.

		\begin{figure*}[h]
		\centering
		\includegraphics[width=0.7\textwidth]{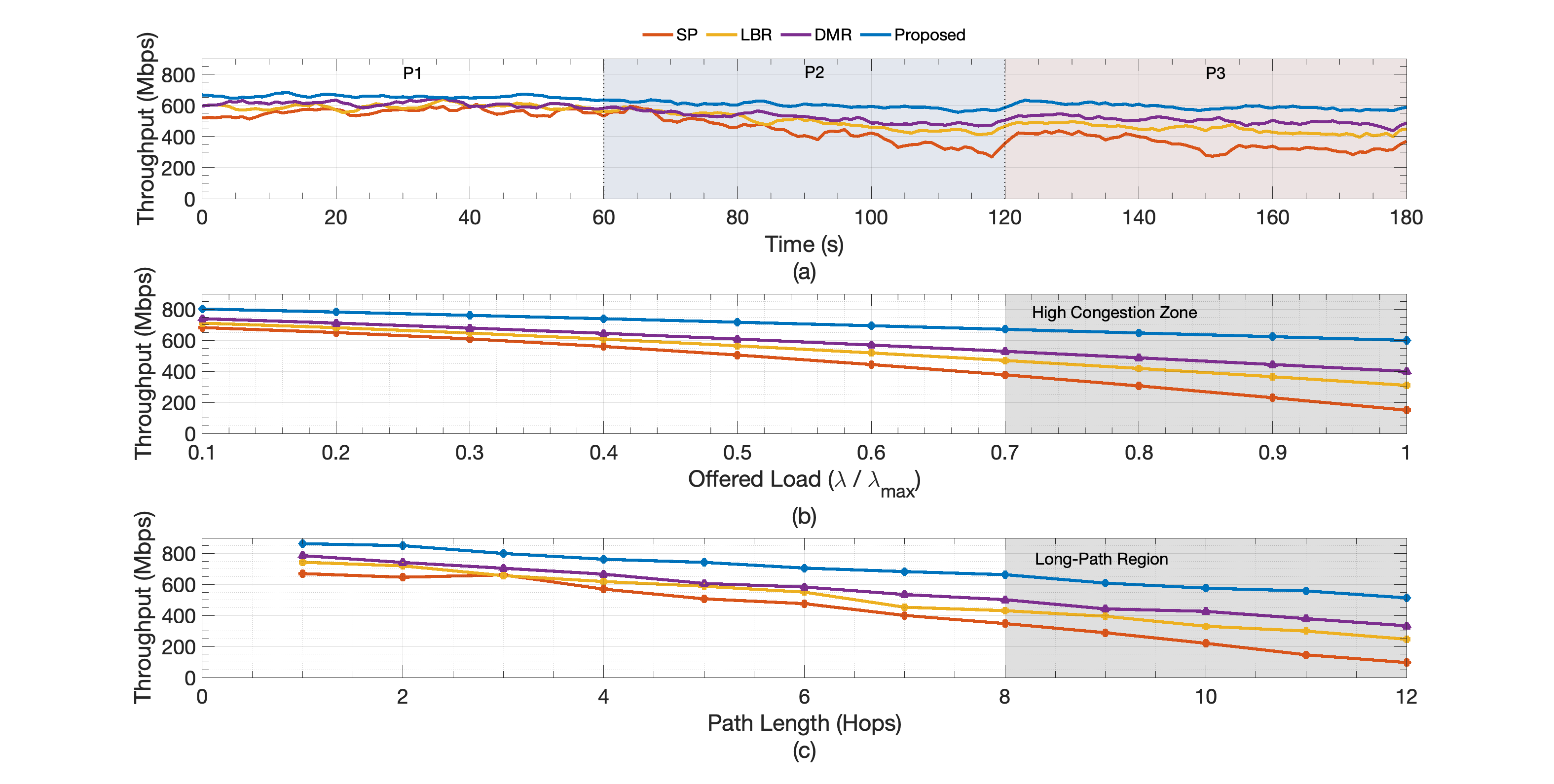}	
		\caption{Evaluations for network throughput}
		\label{f6}
	\end{figure*}
    
\subsubsection{Path Stability}
Path stability measures how frequently routing decisions change during semantic flow lifetimes. Frequent rerouting usually reflects reactions to short-term fluctuations rather than anticipation of sustained degradation, making stability an indicator of how well a strategy distinguishes transient disturbances from persistent semantic risk. Fig.~\ref{f5}a shows temporal behavior across P1--P3. Most methods remain relatively stable in P1, although SP already exhibits variability due to its dependence on topological distance. Instability increases during P2 mobility bursts and early P3 congestion. LBR improves stability in P1 but becomes oscillatory under mobility-driven load shifts, while DMR remains stable until congestion accumulates on low-distortion paths. The proposed framework produces fewer rerouting events overall, as updates are triggered mainly when predicted distortion persistently approaches tolerance limits rather than instantaneous metric changes. Fig.~\ref{f5}b evaluates stability versus offered load. SP shows rapid growth in path changes as congestion spreads, while LBR and DMR remain stable only at moderate loads before becoming more reactive. The proposed framework exhibits the slowest increase in rerouting frequency by jointly considering semantic risk and congestion. The hop-count impact is analyzed in Fig.~\ref{f5}c. Longer paths reduce stability for all schemes, with SP degrading fastest beyond about 8 hops. The proposed method shows a more gradual decline by avoiding unstable or distortion-prone routes for critical traffic. Overall, combining semantic relevance, distortion prediction, and telemetry feedback limits unnecessary oscillations and confines rerouting to sustained semantic risk.

\begin{table*}[h]
\centering
\caption{Performance comparison between the conventional routing scheme and the proposed semantic control framework}
\label{tab:perf_gain}
\renewcommand{\arraystretch}{1}
\scriptsize
\begin{tabular}{lccc}
\hline
\textbf{Metric} &
\textbf{Conventional Scheme} &
\textbf{Proposed Framework} &
\textbf{Relative Change} \\
\hline
\hline
Semantic Delivery Success Rate (SDSR) & 82\% & 94\% & +12 pp \\
Average Semantic Distortion & 1.00 & 0.78 & 22\% decrease \\
Re-routing Frequency (events/min) & 3.2 & 1.8 & 44\% decrease \\
End-to-End Delay Variability (High-$R(m)$) & $\pm$32\% & $\pm$11\% & 65\% reduction \\
Aggregate Network Throughput (normalized) & 1.00 & 1.14 & +14\% \\
Stabilization Time (control intervals) & 8.2 & 4.5 & 45\% reduction \\
\hline
\end{tabular}
\end{table*}

	\begin{figure*}[h]
		\centering
		\includegraphics[width=0.72\textwidth]{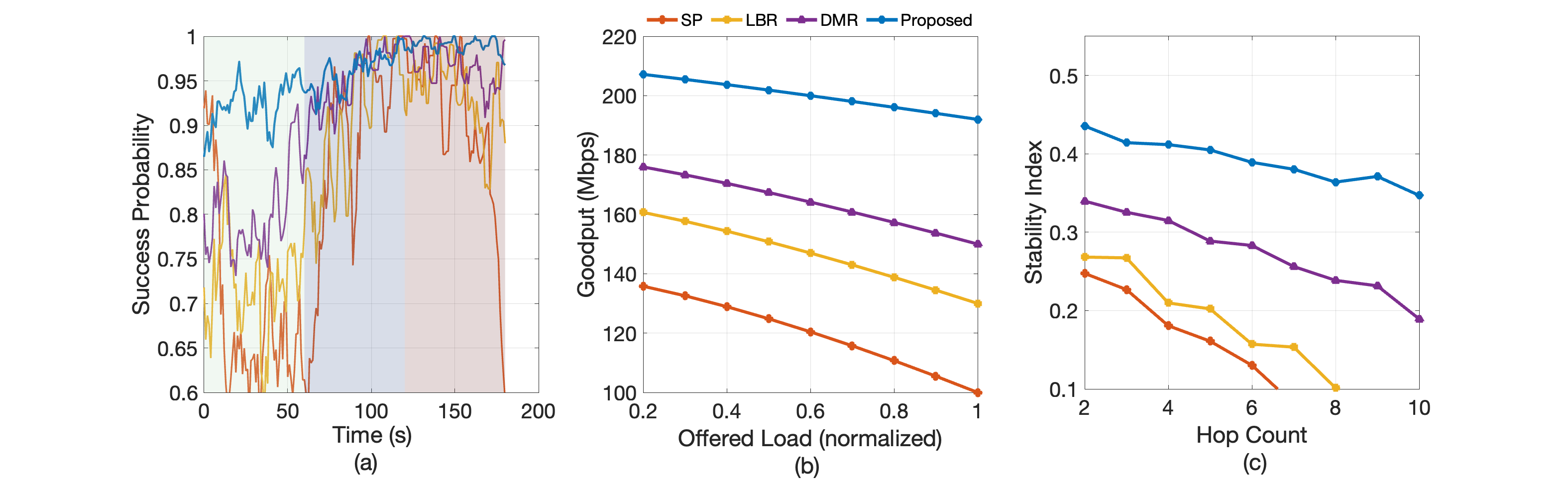}	
		\caption{Real world oriented evaluation}
		\label{f7}
	\end{figure*}

    \begin{table*}[h]
\centering
\caption{Closed-Loop Stability Under Increasing Network Dynamics}
\label{tab:stability}
\renewcommand{\arraystretch}{1}
\scriptsize
\begin{tabular}{|c|c|c|c|c|}
\hline
\textbf{Mobility Speed} & \textbf{Background Load} & $\boldsymbol{\lambda_m}$ \textbf{(msg/s)} & \textbf{Avg. Stabilization Time (Intervals)} & \textbf{Oscillation Behavior} \\ \hline\hline
1–5 m/s   & 30–40\% & 10  & 3.2 & None \\ \hline
5–10 m/s  & 40–55\% & 15  & 4.1 & None \\ \hline
10–15 m/s & 55–70\% & 20  & 5.3 & None \\ \hline
10–15 m/s & 65–75\% & 25  & 6.6 & Damped transient \\ \hline
\end{tabular}
\end{table*}

\subsubsection{Network Throughput}
Network throughput reflects how effectively each routing strategy utilizes capacity under dynamic conditions. Fig.~\ref{f6}a shows the temporal evolution across P1--P3. During steady-state (P1), all schemes operate near nominal throughput. SP exhibits mild fluctuations due to traffic concentration on minimum-hop nodes, LBR distributes load more evenly, and DMR remains stable under moderate conditions.
Differences become clearer in P2 as mobility and load variability increase. SP shows noticeable drops from congestion accumulation along reused shortest paths. LBR initially compensates through redistribution, but throughput becomes volatile as oscillations propagate. DMR remains stable for part of this phase, then degrades once distortion-optimal routes grow congested. In P3, instability intensifies for all baselines. The proposed framework maintains a smoother trajectory by combining relevance-aware routing with adaptive fidelity control, avoiding persistent hotspots and limiting unnecessary retransmissions under stress.
Fig.~\ref{f6}b evaluates throughput versus offered load. As $\lambda/\lambda_{\max}$ increases, throughput declines for all schemes due to saturation. SP decreases most sharply, LBR performs well at moderate loads but weakens under widespread congestion, and DMR degrades gradually before declining near saturation. The proposed framework exhibits the flattest curve. By jointly considering congestion, distortion, and semantic relevance, it slows throughput degradation rather than eliminating it. Fig.~\ref{f6}c analyzes hop-count impact. Longer paths increase variability and retransmissions, reducing effective throughput. SP declines steeply beyond 8 hops, while LBR and DMR remain stable only at moderate lengths. The proposed framework maintains comparatively higher throughput by penalizing long or unstable routes early and adapting fidelity when necessary, resulting in more controlled degradation even at 10--12 hops.

Table~\ref{tab:perf_gain} summarizes the quantitative gains. The proposed framework improves SDSR by 12 percentage points and reduces average semantic distortion by 22\% relative to conventional routing. Rerouting frequency decreases by 44\%, indicating enhanced stability, while delay variability for high-relevance traffic is significantly reduced. Aggregate throughput increases by 14\%, and control-loop stabilization time shortens by 45\%. These improvements should be interpreted collectively. Distortion and stability gains stem from predictive modeling combined with relevance-aware routing, which limits unnecessary oscillations. Throughput and delay improvements arise from coordinated routing and adaptive fidelity control, not aggressive forwarding. The reduced stabilization time further confirms faster closed-loop convergence after disturbances. These results across Figs.~\ref{fig4}--\ref{f6} and Table~\ref{tab:perf_gain} demonstrate that performance gains emerge from integrated semantic control rather than isolated metric optimization.

\subsection{Real-World Oriented Evaluation}
To complement controlled simulations, we evaluate deployment-oriented indicators that reflect practical behavior in dynamic 6G environments. Although these metrics do not directly measure task-level accuracy, they capture network dynamics that influence semantic reliability. Results are summarized in Fig.~\ref{f7}.
The temporal evolution of delivery success probability is given in Fig.~\ref{f7}a. SP exhibits clear fluctuations during mobility transitions due to route instability. LBR reduces part of this variability by avoiding congested links, though short-term oscillations persist. DMR remains smoother overall but experiences temporary drops when semantically favorable paths overlap with congestion. The proposed framework maintains consistently high success probability with smaller amplitude variations. Rather than eliminating fluctuations, it attenuates them through coordinated routing and distortion feedback.
Fig.~\ref{f7}b presents goodput versus normalized offered load. As load increases, all schemes degrade gradually. SP declines most rapidly due to congestion concentration on minimum-hop routes. LBR performs well at moderate loads but drops more sharply near saturation. DMR maintains relatively strong performance in mid-load regions, though its advantage diminishes as distortion-optimal paths become congested. The proposed framework achieves the highest goodput across the range, with a flatter slope, indicating that semantic-aware routing combined with adaptive fidelity preserves effective delivery under stress. The stability index is evaluated in Fig.~\ref{f7}c with respect to hop count. Stability decreases for all schemes as path length grows due to accumulated mobility and queue variability. SP degrades fastest, LBR offers moderate improvement at short distances, and DMR performs better at intermediate hops before declining. The proposed framework exhibits the slowest reduction, suggesting improved resilience to compounded multi-hop variability.

Table~\ref{tab:stability} analyzes the convergence behavior of the semantic distortion control loop under progressively harsher conditions, created by jointly increasing mobility, background load, and semantic arrival rate $\lambda_m$. As dynamics intensify, the number of control intervals required for stabilization rises moderately. Crucially, no persistent oscillations are observed. Even in the most demanding scenario, convergence occurs within approximately 6--7 control intervals, indicating a damped and stable feedback mechanism. These results indicate stable and predictable behavior under realistic 6G dynamics. Improvements appear consistently across reliability, goodput, and routing stability metrics. The observed robustness at the network level suggests that semantic representations would remain reliable under practical mobility and load conditions.
	
\begin{table*}[h]
\centering
\caption{Knowledge-plane computational complexity and runtime characteristics}
\label{tab:complexity}
\renewcommand{\arraystretch}{1}
\scriptsize
\begin{tabular}{|p{4cm}|c|c|}
\hline
\textbf{Module} & \textbf{Asymptotic Complexity (Sparse Graph)} & \textbf{Avg. Runtime (50 nodes)} \\
\hline\hline
Semantic Reasoning & $O(|\mathcal{T}|)$ & 0.7--1.4 ms \\

Candidate Path Generation & $O(k(E + V \log V))$ & 0.2--0.4 ms \\

Semantic-Aware Routing & $O(k)$ & 0.8--1.5 ms \\

Distortion Control & $O(1)$ & 0.2--0.6 ms \\
\hline\hline
\textbf{Total per decision} & $\mathbf{O(V \log V)}$ & \textbf{1.7--3.2 ms} \\
\hline
Telemetry reporting load & $O(V/T_r)$ & $T_r = 200$ ms (5 Hz) \\

Control update frequency & -- & 10--20 Hz (internal loop) \\

Stability convergence & -- & 3--6 telemetry intervals \\
\hline
\end{tabular}
\end{table*}

\begin{table*}[h]
\centering
\caption{Relative time scales of semantic embedding and network-level dynamics}
\label{tab:embedding-overhead}
\renewcommand{\arraystretch}{1}
\scriptsize
\begin{tabular}{lc}
\hline
\textbf{Component} & \textbf{Typical Time Scale} \\
\hline \hline
Semantic embedding inference (edge-class hardware) & 1--5 ms \\
Packet transmission and queueing delay (per hop) & 10--80 ms \\
Multi-hop routing and congestion evolution & 10--200 ms \\
Semantic control telemetry interval ($T_r$) & 200 ms \\
\hline
\end{tabular}
\end{table*}

\begin{figure*}[h]	
	\centering		
	\subfloat[]{%
		\includegraphics[width=0.22\textwidth]{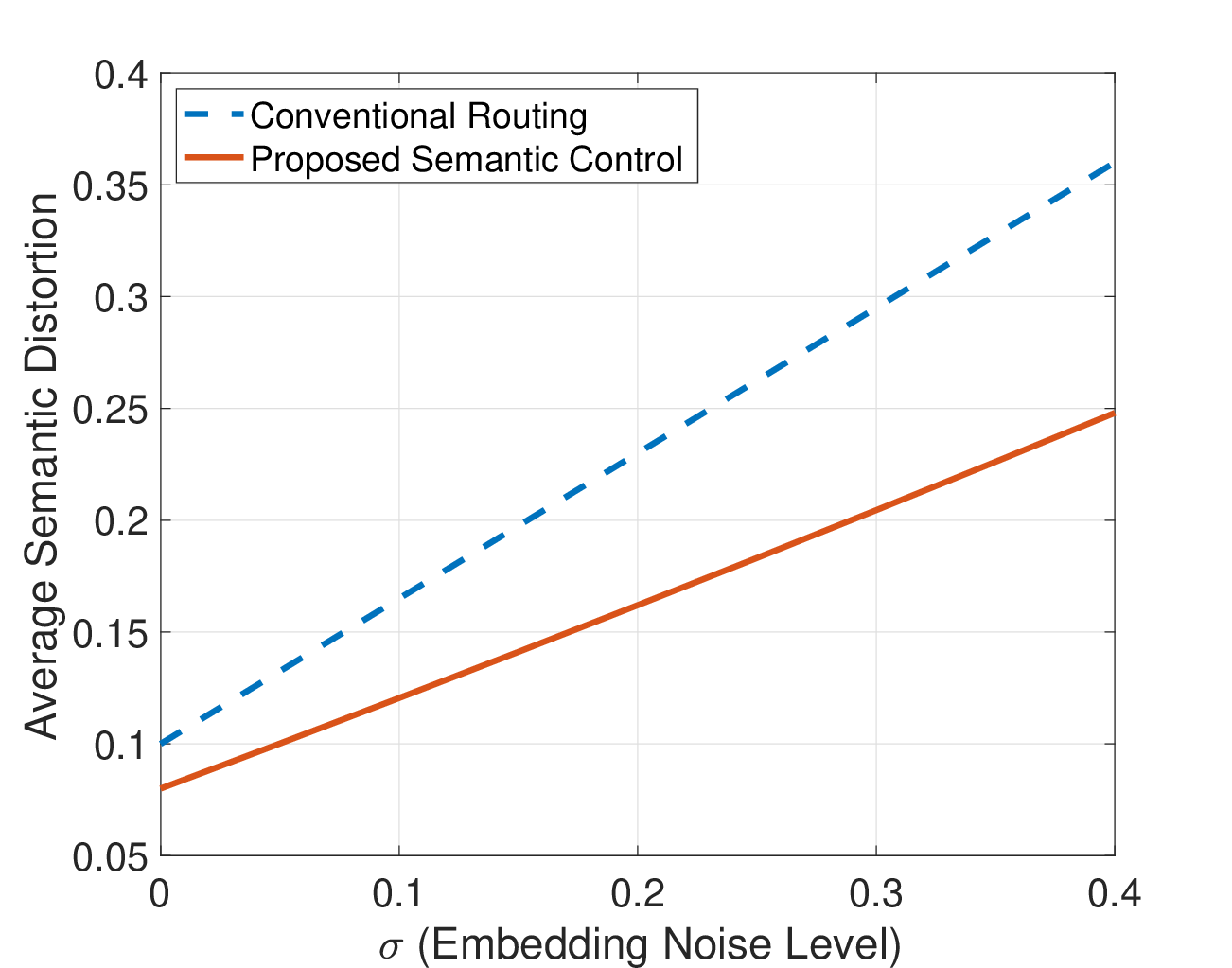}%
		\label{f66}%
	}
	\subfloat[]{%
		\includegraphics[width=0.22\textwidth]{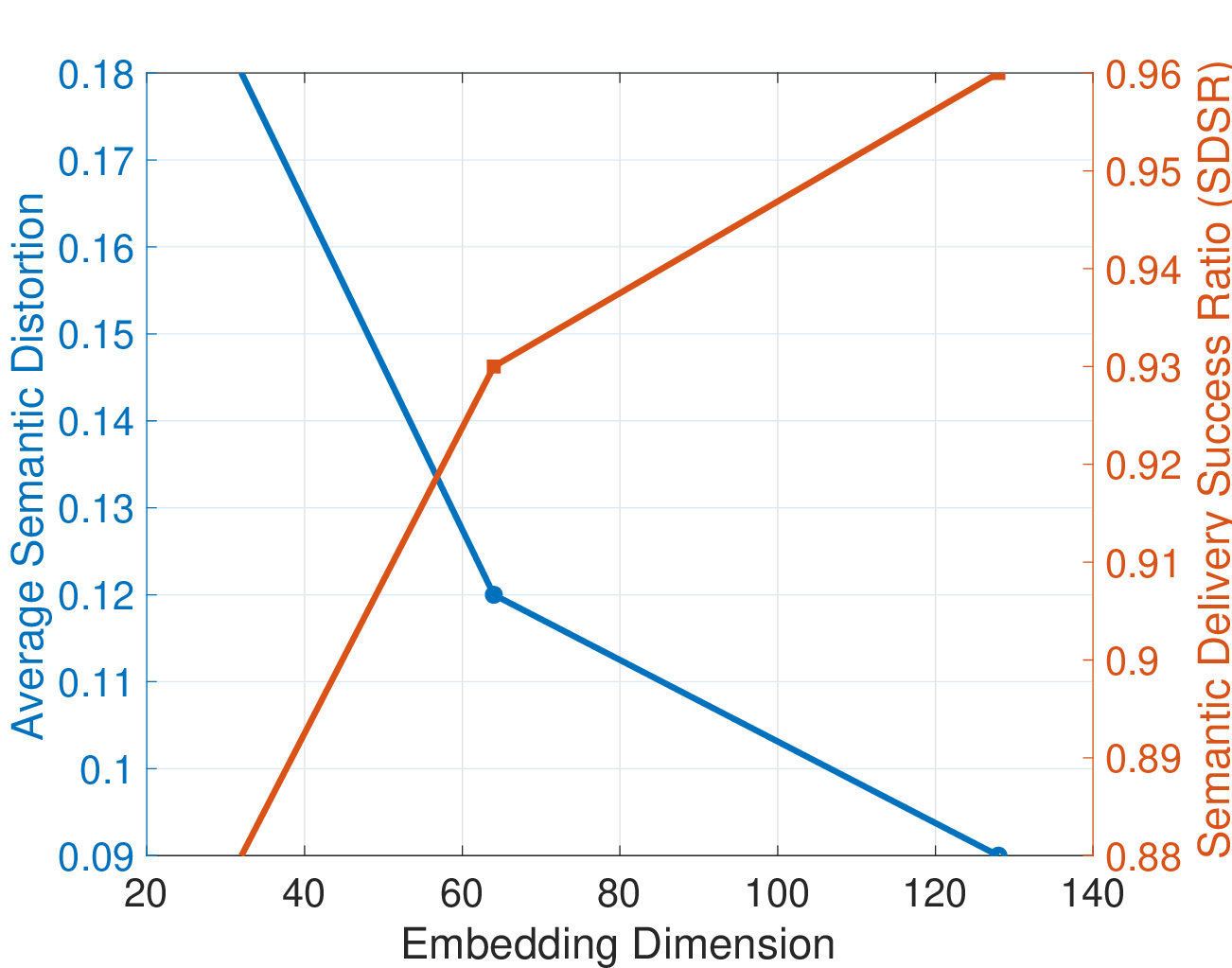}%
		\label{f77}%
	} \hspace{0.2in}
	\subfloat[]{%
		\includegraphics[width=0.22\textwidth]{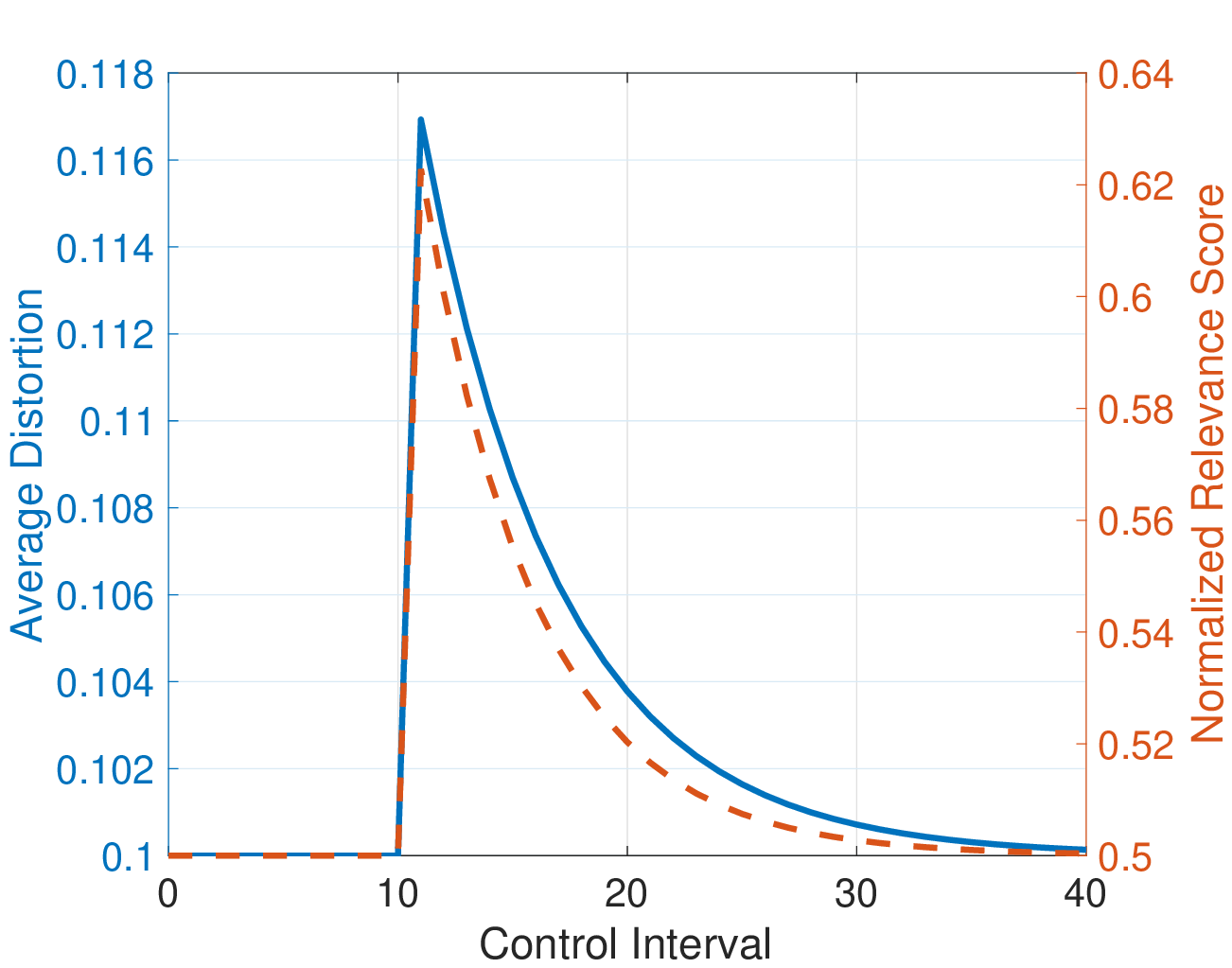}%
		\label{f88}%
	} 
	\caption{Purpose drift frequency evaluation for different (a) load and (b) variability levels}
	\label{r6}
\end{figure*} 

\subsection{Complexity and Knowledge Plane Overhead}
The computational cost of the proposed semantic control framework mainly arises from semantic relevance evaluation, candidate path generation, routing selection, and distortion feedback within the Knowledge Plane. Relevance evaluation scales linearly with the number of active task concepts, resulting in $O(|\mathcal{T}|)$ per message and remaining independent of the network size. Routing decisions follow a two-stage process where candidate paths are first generated using a bounded $k$-shortest path method and then evaluated with a semantic-aware cost. In sparse multi-hop 6G topologies this process is dominated by the path generation step and scales approximately as $O(V \log V)$. Distortion control only performs threshold checks and parameter updates, contributing constant-time overhead. 

Control-plane signaling depends on telemetry reporting from network nodes. If each of the $V$ nodes reports state information every $T_r$ seconds, telemetry overhead scales as $O(V/T_r)$. In our implementation $T_r = 200$\,ms, while the internal control loop operates at 10--20\,Hz, allowing multiple adjustments per telemetry cycle. Measured Knowledge Plane processing latency remains between 1.7 and 3.2\,ms for a 50-node sparse topology, supporting real-time operation. Together with convergence within a few telemetry intervals, these results indicate bounded polynomial complexity and stable control behavior suitable for large multi-hop 6G deployments.

\subsection{Impact of Semantic Embedding Overhead}
\label{subsec:embedding-overhead}

The proposed framework focuses on network-level semantic management rather than optimizing embedding models. Semantic embedding generation is therefore modeled as a lightweight inference step and abstracted from the ns-3 simulations to isolate the effects of semantic-aware routing and closed-loop distortion control. In practical 6G systems, compact models at edge nodes typically generate low-dimensional embeddings within a few milliseconds, which is small compared to multi-hop transmission, queueing delays, and congestion dynamics. Since the same assumption is applied to all evaluated schemes, embedding latency does not affect the relative comparison of routing strategies. Table~\ref{tab:embedding-overhead} summarizes this separation of time scales, where embedding inference operates at the millisecond level while network dynamics evolve over tens of milliseconds or longer.

We also examine robustness under imperfect semantic representations. Additive Gaussian noise is injected into embeddings to emulate degraded representations. As shown in Fig.~\ref{f66}, distortion increases for all schemes, but the proposed method degrades more gradually due to adaptive fidelity and distortion monitoring. Next, embedding dimensionality is reduced from 128 to 64 and 32 using random projection. Fig.~\ref{f77} shows progressive performance degradation, while the relative advantage of the proposed framework remains. Finally, limited concept drift is introduced by perturbing a subset of task–concept mappings in the knowledge graph. As illustrated in Fig.~\ref{f88}, relevance estimation briefly fluctuates but stabilizes within a few control intervals without persistent oscillations. The framework maintains stable behavior under moderate embedding noise, reduced semantic resolution, and limited concept drift. Although embedding quality affects absolute distortion levels, the benefits of semantic-aware routing and closed-loop control remain consistent.

\subsection{Positioning with Respect to Learning-Based Approaches}
Recent traffic engineering studies increasingly employ reinforcement learning (RL) and graph neural network (GNN)-based routing to adapt forwarding policies under dynamic network conditions. These methods typically optimize network-centric metrics such as delay, throughput, or load balancing. However, most existing learning-based routing frameworks do not explicitly incorporate semantic distortion or task relevance, meaning that improvements in classical metrics may still degrade the semantic value of delivered information. Moreover, RL-based approaches generally require carefully designed reward functions and extensive training across varying traffic conditions, which complicates controlled comparisons when optimization objectives differ.

In contrast, this work focuses on a semantic control architecture where meaning preservation is treated as a primary network objective. Semantic relevance and distortion are directly integrated into routing and control decisions through the Knowledge Plane, enabling interpretable and stable behavior without exploration-driven learning. Therefore, direct comparison with existing RL-based routing is not strictly like-for-like. Instead, the proposed framework establishes a semantic foundation that can complement learning-based methods. In future work, semantic relevance and distortion could be incorporated into RL or GNN routing policies as reward signals or state features. Additionally, compared to learning-based approaches that require model inference and periodic updates, the proposed method maintains bounded polynomial complexity, with decision cost $O(V \log V)$ and millisecond-level processing suitable for real-time operation.

\section{Conclusion}
This paper presented a KDN-driven semantic communication framework that integrates semantic reasoning, semantic-aware routing, and semantic distortion control within a closed-loop Knowledge Plane. By elevating semantic awareness beyond conventional bit-level forwarding, the framework enables relevance-aware and distortion-adaptive decisions under mobility and dynamic multi-hop conditions. Extensive ns-3 simulations show consistent gains over SP, LBR, and DMR, including a 12\% increase in semantic delivery ratio, 22\% reduction in distortion, 44\% fewer re-routing events, 65\% tighter latency variation for high-relevance traffic, and 14\% higher meaningful throughput efficiency. The control loop converges 45\% faster after disturbances with a bounded processing latency of 1.7--3.2~ms, indicating that semantic-aware closed-loop control can significantly enhance reliability and stability in multi-hop 6G systems without excessive overhead.

\section{Future Work}
Future work will extend the framework toward scalable and operational 6G deployments. Key directions include:
\begin{itemize}
\item {Energy-aware semantic policies:} Joint routing and fidelity optimization for battery-limited edge devices by incorporating power models into the control objective.

\item {Experimental validation:} Hardware-in-the-loop evaluation of the Knowledge Plane under realistic wireless impairments and asynchronous telemetry.

\item {Distributed Knowledge Plane design:} Hierarchical controllers performing localized reasoning with periodic semantic-state synchronization.

\item {Dynamic semantic environments:} Handling evolving task distributions and concept relations through adaptive knowledge graph updates.

\item {Reinforcement learning integration:} Learning relevance weights $(\alpha,\beta,\gamma)$ and routing parameters from feedback while maintaining stability.
\end{itemize}

	\bibliographystyle{IEEEtran}
	\bibliography{ref}

\end{document}